\documentclass[fleqn,usenatbib]{mnras}
\usepackage{newtxtext,newtxmath, longtable}

\usepackage[T1]{fontenc}

\DeclareRobustCommand{\VAN}[3]{#2}
\let\VANthebibliography\thebibliography
\def\thebibliography{\DeclareRobustCommand{\VAN}[3]{##3}\VANthebibliography}

\usepackage{graphicx}	
\usepackage{amsmath}	
 \usepackage{longtable}
 \usepackage{xtab,afterpage}

 \usepackage{xcolor}
 \usepackage{xspace}
\newif\ifhighlighton
\highlightonfalse   

\DeclareUnicodeCharacter{2212}{-}

\title[Anomaly Detection for Radio Transients]{Finding radio transients with anomaly detection and active learning based on volunteer classifications}

\author[A. Andersson et al.]{Alex Andersson,$^{1}$\thanks{E-mail: alexander.andersson@physics.ox.ac.uk}
Chris Lintott,$^{1}$
Rob Fender,$^{1,2}$
Michelle Lochner,$^{3,4}$
Patrick Woudt,$^{2}$
Jakob van den Eijnden,$^{5,6}$
\newauthor{Alexander van der Horst,$^{7}$
Assaf Horesh,$^{8}$
Payaswini Saikia,$^{9}$
Gregory R. Sivakoff,$^{10}$ 
Lilia Tremou,$^{11}$
}
\newauthor{Mattia Vaccari$^{12, 13, 14}$
}
\\
$^{1}$Astrophysics, Department of Physics, University of Oxford, Denys Wilkinson Building, Keble Road, Oxford OX1 3RH, UK\\
$^{2}$Department of Astronomy, University of Cape Town, Private Bag X3, Rondebosch 7701, South Africa\\
$^{3}$Department of Physics and Astronomy, University of the Western Cape, Bellville, Cape Town 7535, South Africa\\
$^{4}$South African Radio Astronomy Observatory, 2 Fir Street, Black River Park, Observatory, Cape Town 7925, South Africa\\
$^{5}$Department of Physics, University of Warwick, Coventry CV4 7AL, UK\\
$^{6}$Anton Pannekoek Institute for Astronomy, Universiteit van Amsterdam, Science Park 904, 1098, XH, Amsterdam, The Netherlands\\
$^{7}$Department of Physics, The George Washington University, 725 21st Street NW, Washington, DC 20052, USA\\
$^{8}$Racah Institute of Physics, The Hebrew University of Jerusalem, Jerusalem 91904, Israel\\
$^{9}$Center for Astro, Particle and Planetary Physics, New York University Abu Dhabi, PO Box 129188 Abu Dhabi, UAE\\
$^{10}$Department of Physics, University of Alberta, CCIS 4-181, Edmonton AB, T6G 2E1, Canada\\
$^{11}$National Radio Astronomy Observatory, Socorro, NM 87801, USA\\
$^{12}$Inter-University Institute for Data Intensive Astronomy, Department of Astronomy, University of Cape Town, 7701 Rondebosch, Cape Town, South Africa\\
$^{13}$Inter-University Institute for Data Intensive Astronomy, Department of Physics and Astronomy, University of the Western Cape, Bellville 7535, Cape Town, South Africa\\
$^{14}$INAF - Istituto di Radioastronomia, via Gobetti 101, 40129 Bologna, Italy
}

\date{Accepted XXX. Received YYY; in original form ZZZ}

\pubyear{2025}

\begin{document}
\label{firstpage}
\pagerange{\pageref{firstpage}--\pageref{lastpage}} 
\maketitle

\begin{abstract}
In this work we explore the applicability of unsupervised machine learning algorithms to finding radio transients. Facilities such as the Square Kilometre Array (SKA) will provide huge volumes of data in which to detect rare transients; the challenge for astronomers is how to find them. We demonstrate the effectiveness of anomaly detection algorithms using 1.3~GHz light curves from the SKA precursor MeerKAT. We make use of three sets of descriptive parameters (‘feature sets’) as applied to two anomaly detection techniques in the \textsc{Astronomaly} package and analyse our performance by comparison with citizen science labels on the same dataset. Using transients found by volunteers as our ground truth, we demonstrate that anomaly detection techniques can recall over half of the radio transients in the 10 per cent of the data with the highest anomaly scores. We find that the choice of anomaly detection algorithm makes a minor difference, but that feature set choice is crucial, especially when considering available resources for human inspection and/or follow-up. Active learning, where human labels are given for just 2 per~cent of the data, improves recall by up to 20 percentage points, depending on the combination of features and model used. The best performing results produce a factor of 5 times fewer sources requiring vetting by experts. This is the first effort to apply anomaly detection techniques to finding radio transients and shows great promise for application to other datasets, and as a real-time transient detection system for upcoming large surveys.
\end{abstract}




\section{Introduction}
\label{sec:intro}

The Square Kilometre Array (SKA; \citealp{Braun2019arXiv191212699B}) promises to deliver unprecedented data across many areas of astrophysics. Already, SKA precursors such as ASKAP \citep{Hotan2021} and MeerKAT \citep{Jonas2016} are providing a unique combination of highly sensitive observations taken at regular cadence over a wide field-of-view. These observations then allow for untargeted searches for radio transients in surveys such as ThunderKAT \citep{Fender2016} and VAST (Variables and Slow Transients, \citealp{Murphy2013}). ThunderKAT observed known radio transients (e.g. X-ray binaries; XRBs), but the unique capabilities of MeerKAT have allowed for serendipitous discovery of magnetically active and flaring stars (\citealp{Driessen2020, Driessen2021, Andersson2022, Chastain2023, Chastiain2025, Peters2023, Fijma2024, Smirnov2025}, as well as Mlangeni et al. and Nyamai et al., \textit{in prep}), scintillating AGN and unique pulsar behaviour \citep{Driessen2022, Rowlinson2022}. The VAST survey's wide-field imaging with ASKAP has discovered large, scintillation-inducing plasma arcs \citep{YWang2021}, mysterious transients near the Galactic centre \citep{ZWang2021, ZWang2022}, and many highly variable systems on minute timescales \citep{YWang2023}. Furthermore, recent observations at timescales of a few minutes have led to uniquely long-period transients\footnote{Here, minute-timescales are considered `long' with respect to typical pulsar periods of $<1$~s.}, pulsars and white dwarf systems \citep{Hurley-Walker2022, Hurley-Walker2023, HurleyWalker2024, Caleb2022, Caleb2024NatAs...8.1159C, Pelisoli2023, deRuiter2024arXiv240811536D}. Finally, imaging on very short timescales can immediately localise fast radio bursts (FRBs) to high precision \citep{Rhodes2023, Driessen2024}. The combination of all aspects of a dataset - its spectral coverage, time baseline and polarimetry - has been shown to provide a unique discovery space for pulsars and transients \citep{Heywood2023, Smirnov2024}.

Despite this interesting zoo of sources, radio transients and variables in observations around 1 GHz are relatively rare, of order 1-4 per cent of sources per observational field \citep{Mooley2016,Murphy2017,Sarbadhicary2021}. At the same time, current and upcoming radio facilities promise to overwhelm astronomers with data, creating a needle-in-a-haystack problem of detecting the rare and interesting transients in a deluge of high-cadence, highly sensitive observations. As a demonstration of this, current deep imaging with MeerKAT can detect $\sim6000$ sources/deg$^2$ \citep{Heywood2022}. The SKA-Mid Array, when complete, will be able to reach the same depth in approximately 12hrs\footnote{Calculated from \url{https://sensitivity-calculator.skao.int/mid}}.
If the array were to conduct non-overlapping 12hr pointings over an entire year, it would observe upwards of 4,000,000 sources (over less than 1/40\textsuperscript{th} of the sky).
If time-variable phenomena make up approximately 2\% of all sources in a given radio image at any sensitivity limit, this means that the SKA-Mid will see up to $\sim40,000$ transients and variables in its first year of operation. 
There is therefore a need to develop and test a range of methods in order to discover transients in data-streams, with the enormous data-rates prompting the need for low-latency searches in order to avoid storing unreasonably large quantities of data. An example of this can be seen in the case of the MeerTRAP \citep[More TRansients And Pulsars,][]{Stappers2016} backend employed on MeerKAT. The MeerTRAP user supplied equipment taps into the MeerKAT datastream, commensally searching for fast transients such as FRBs and bright single pulses from pulsars. This effort requires real-time candidate processing and filtering in order to reduce the data rate to reasonable volumes, with final candidates requiring additional rapid confirmation or rejection by the team in order to not overwhelm storage capacity \citep{Malenta2020, Jankowski2022}.

This data-intensive regime is not unlike the problem that faces optical astronomers, who have spent several years preparing for the Vera C. Rubin Observatory to come online and deliver the Legacy Survey of Space and Time (LSST), consisting of millions of transient alerts per night in a cumulative volume of petabyte scale data \citep{Ivezic2019}. One proposed solution to this problem has been into anomaly detection methods - these are unsupervised (that is, untrained) machine learning models that seek to separate rare anomalies from everything else. Much of this work has been done relating to supernovae, due to their relevance to a wide range of astrophysics and cosmology but the scarcity of spectroscopic follow-up. For example, \citet{Pruzhinskaya2019} and \citet{Malanchev2021} demonstrate the applicability of isolation forests and local outlier factor algorithms (see  section \ref{sec:anomaly}) to detecting rare supernovae classes and novel transients. 
It is not only the models themselves that have been tested, but also the descriptions of the input data. As mentioned in section \ref{sec:features}, the choice of features - attributes or parameters that are calculated or learnt from raw data, which are then fed into downstream tasks - is paramount in machine learning. These features might be extracted based on some previously known importance (e.g. the peak magnitude of a supernova) or, for large and complex datasets, learnt from the data themselves (deep learning). For example, \citet{Villar2021} use an autoencoder framework to extract salient features from light curve simulations directly, as opposed to using previously known metrics. However, deep learning architectures are not always the answer - for example, \citet{Muthukrishna2021} demonstrate that their overflexibility can make them less suitable for detection of anomalies in real time when compared to parametric models.

One potential pitfall of these anomaly detection techniques is that anomalies identified by a specific algorithm may not match what the end-user (i.e. a science team) deem to be interesting. By incorporating domain knowledge into machine learning models to refine their performance one can, for example, separate anomalous instrumental effects from bona fide astrophysical phenomena. This is known as active learning, the core idea of which is that an algorithm can query a human annotator (sometimes called an "oracle") to label a small subset of data from which the learner can train. If the learner can choose the data for the annotator to label, the hope is that performance will increase by labelling only a small but informative subset of the data.
This active learning for anomaly detection has been tested on simulated data in preparation for LSST and real data streams from the Zwicky Transient Facility and the Open Supernova Catalog \citep{Ishida2021, Pruzhinskaya2023}.

Despite the advances in appropriate feature use, anomaly detection and active learning in optical surveys, these tools have yet to be applied to the rapidly-growing field of wide-field radio transient surveys, with the most similar use-cases being for investigating anomalous spectrograms and parameter optimisation in low-frequency radio data and pipelines \citep{Rowlinson2019A&C....27..111R, Mesarcik2023A&A...680A..74M}. The \textsc{Astronomaly} package \citep[][]{Lochner2021} provides a general framework for discovering anomalies in astronomical datasets, which we can use to test these different aspects of the discovery process. \textsc{Astronomaly} has three main scientific steps consisting of feature extraction, anomaly detection and active learning, whilst also providing data management, pre-processing and visualisation tools. Exploration of how these different stages perform, with different features and algorithms applied to a novel dataset form sections \ref{sec:features}--\ref{sec:active} of this work, though we note that they are all built into \textsc{Astronomaly}'s framework.  \textsc{Astronomaly} has been used to discover variable and flaring stars in optical surveys \citep{Webb2020, Webb2021}, as well as rare and unique morphologies of optical galaxies, mergers and radio sources \citep{Etsebeth2023, Mohale2023, Lochner2023}.

In this paper we describe the first exploration of applying unsupervised anomaly detection techniques to large samples of light curves from radio telescopes. This is motivated by both the ever-growing data rates from modern radio telescopes as discussed earlier, along with the need for robust methods for finding transients that improve upon previous searches. Previous studies \citep[such as][]{Driessen2022,Andersson2023} have shown that many variables and transients are difficult to identify with simple statistics (see equations \ref{eq:eta} and \ref{eq:V}). As we are interested in finding such transients in large datasets (e.g. from the SKA), these findings were part of the motivation for studying and applying anomaly detection techniques.

We make use of 8874 light curves from the MeerKAT telescope as part of commensal ThunderKAT observations, analysed by citizen scientists in search of variables and transients \citep{Andersson2023}. By its nature, unsupervised learning requires no classifications and therefore easily scales to large datasets as there is no need for costly and/or time-consuming labelling of data nor the training of large models. By the same token, however, it is hard to verify performance of unsupervised learning methods without some known ‘ground truth' (see e.g. \citealp{Giles2019}). Therefore, we make use of the citizen science classifications as ground truth against which to compare anomaly detection methods from the \textsc{Astronomaly} package. 
We describe the MeerKAT and citizen science data in section \ref{data} before detailing the features used in this work in section \ref{sec:features}. We then demonstrate how standard anomaly detection performs on these data in section \ref{sec:anomaly} and how the active learning built into \textsc{Astronomaly} improves this performance in section \ref{sec:active}. We discuss this work and its implications before concluding in sections \ref{sec:disc} and \ref{sec:conclude}.


\section{Data: ThunderKAT Observations and Citizen Science}
\label{data}

\defcitealias{Andersson2023}{A23}

This work builds on the data collected and presented in \citet[][A23 hereafter]{Andersson2023}. Here we summarise the observations and resultant data products (light curves), as well as the citizen science classifications that followed. For full details, see sections 2 and 3 of \citetalias{Andersson2023}.

\subsection{Observational Data}
\label{sec:obs}
Our observational data consist of $\sim$500 images taken by MeerKAT from 2018 - 2022 at 1.3 GHz, distributed across 11 fields of the sky and centred on XRBs of interest to the ThunderKAT programme. These observations, typically taken at weekly cadence, consist of scans of a primary calibrator and phase calibrator bookending a 15 minute observation on-source. The observations were reduced in a standard procedure, using the \texttt{OxKAT} scripts \citep{Heywood2020} and their underlying packages to flag, calibrate and image the data. The resultant set of images is heterogeneous in number of visits per field and this heterogeneity is entirely determined by the commensal nature of the observations - if an XRB faded into quiescence it was dropped from the weekly schedule. As a result some fields have only two months worth of observations, whilst fields around bright and active systems such as MAXI~J1820+070, GRS~1915+105 and GX339$-$4 have over 50 visits. 

\begin{figure}
    \includegraphics[width=\columnwidth]{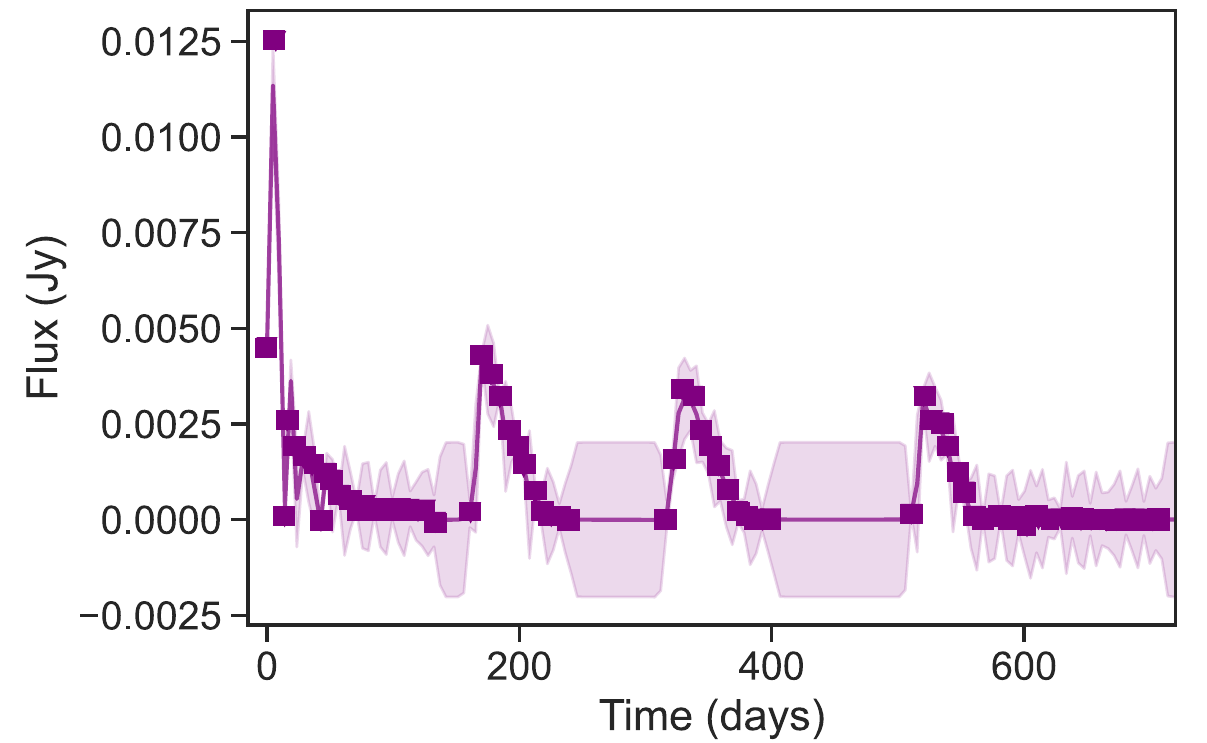}
        \includegraphics[width=\columnwidth]{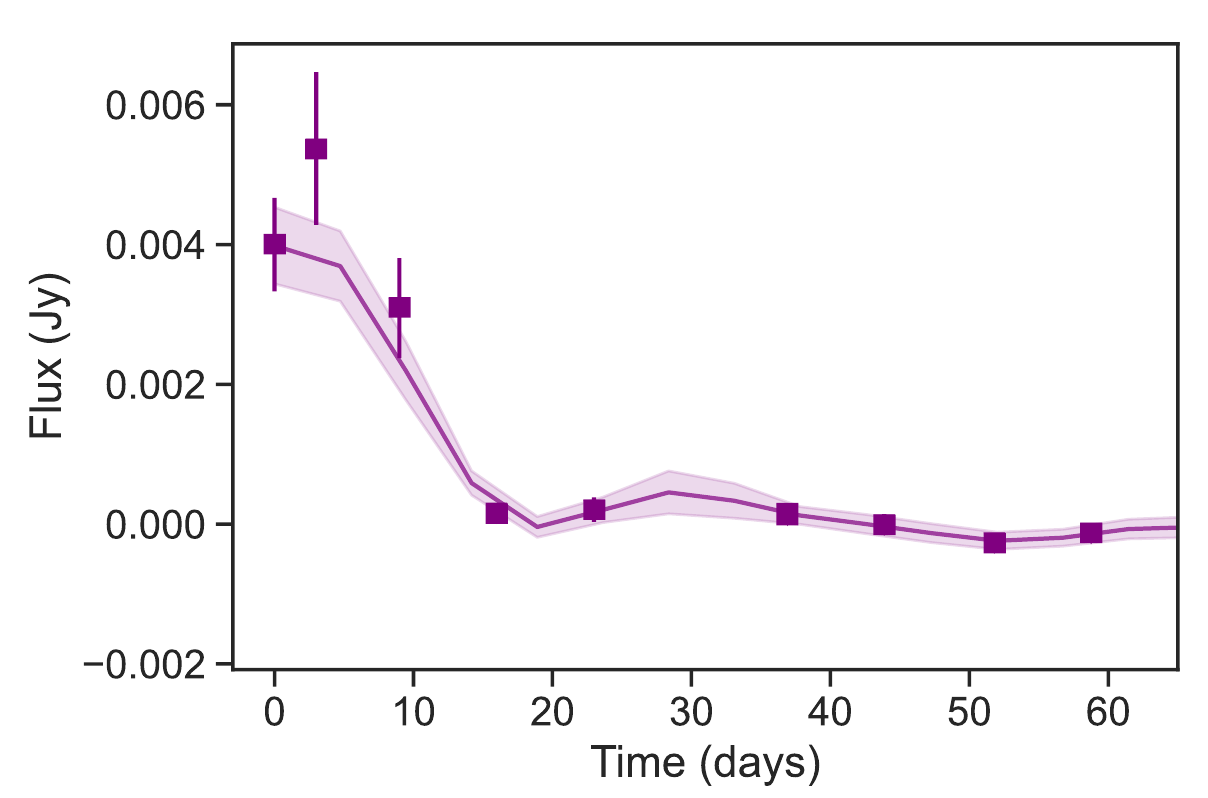}
    \includegraphics[width=\columnwidth]{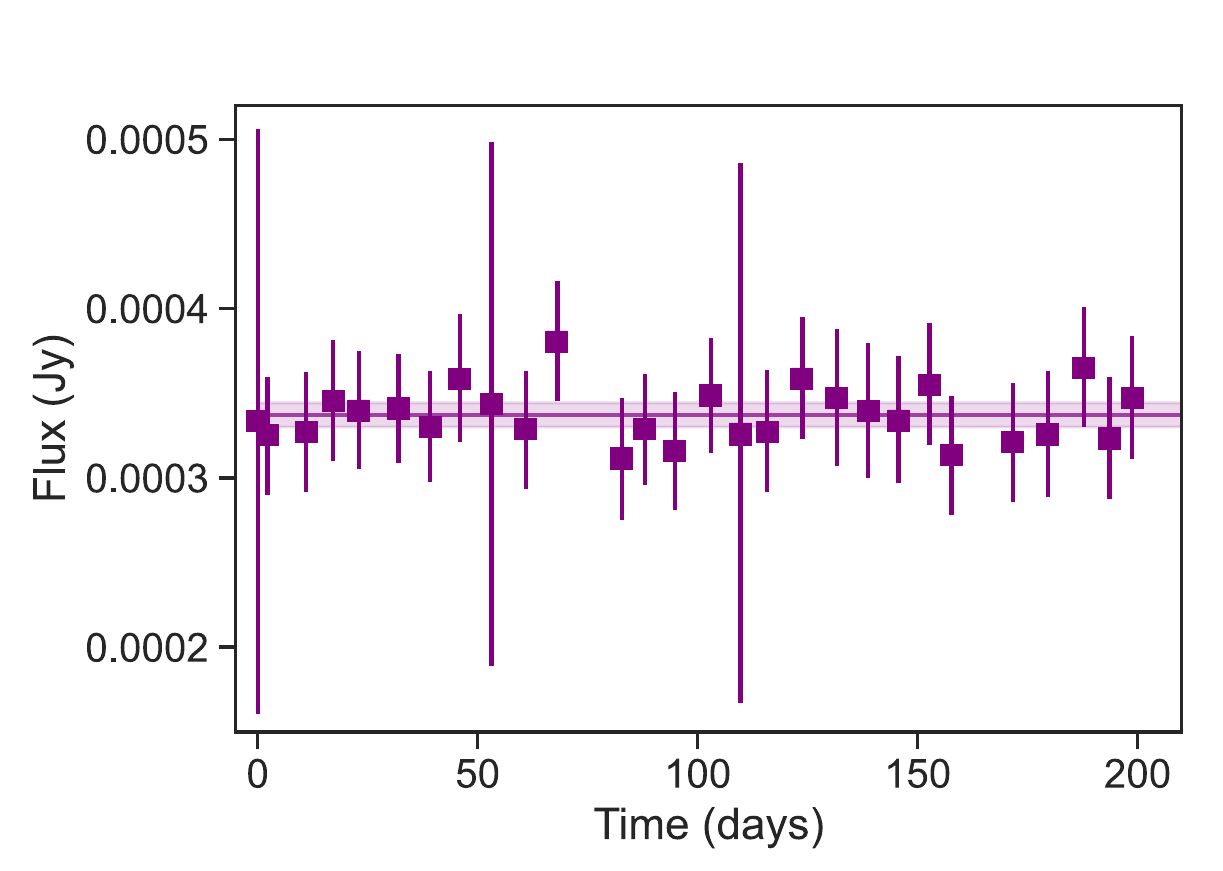}
    \caption{Example light curves of an XRB (MAXI J1820+070, top\protect\footnotemark), a false positive caused by bad data (middle) and a stable light curve (bottom). Our aim is for unsupervised machine learning algorithms to preferentially select light curves similar to the foremost over the latter two. The line and shaded region of each panel denote the mean and 1$\sigma$ uncertainty of the computed Gaussian Process regression, detailed in section \ref{sec:wavelets}.}
    \label{fig:light_curves}
\end{figure}

Once imaged, light curves are extracted from the observations using the Transient Pipeline \citep[\textsc{TraP},][]{Swinbank2015}. The \textsc{TraP} extracts flux density measurements for all sources detected with a S/N $\geq$8 within 1.5$\times$ the primary beam radius of each observation. Each flux measurement comes from a Gaussian fit to each source, fixed to be the size of the PSF in that observation, as we are only interested in point sources and a changing PSF between epochs can induce false variability. The TraP is run such that if, in any observation, a source is found, that position is checked and force-fit in all subsequent epochs (but not in prior images), providing upper-limits in images where the source is absent (e.g. a transient). Examples of a bona fide transient, a false positive and a stable light curve are all visible in Figure \ref{fig:light_curves}, demonstrating the diversity of our data.

\footnotetext{See \cite{Bright2020NatAs...4..697B} for more on the radio behaviour of this source and Bright et al. (\textit{in prep.}) for analysis of the long-term light curve.}

From running the TraP we produce 8874 light curves in total from the 11 XRB fields. As with the images, the number of light curves per field is not uniform, with some fields having many more than others. This is primarily governed by some fields having more diffuse emission, higher noise floors and/or worse \textit{u,v} coverage.

\subsection{Citizen Science Data}

The 8874 light curves described in the previous section were shown to volunteers on the \textit{Bursts from Space: MeerKAT} (BfS:MKT) citizen science project, hosted on the Zooniverse platform \citepalias{Andersson2023}\footnote{\url{https://www.zooniverse.org/projects/alex-andersson/bursts-from-space-meerkat}}. 10 volunteers look at each subject on the platform and determine whether the source in question is truly a transient/variable or a member of a different class - stable sources with flat light curves, imaging artefacts that can appear as spurious transients and so on. Once collected, these 10 votes were aggregated to give a transient vote fraction and one of us (AA) manually inspected all 381 sources where 4/10 or more volunteers said a subject was a transient or variable. From this final selection we have 168 transients and variables in our sample, consisting of 142 newly identified sources and 26 that had been previously studied.

For the purposes of this paper, we will treat the 168 variables and transients found by our volunteers and  verified by us as our ground truth sample. From hereon, any reference to transients refers to this sample, representing $\sim$2 per cent of our light curves. Due to this large imbalance between transients and non-transients we choose to apply anomaly detection methods in order to understand how detectable these rare systems are \textit{without} the need for citizen science verification (e.g. in a low-latency transient detection system). As a note, we do not distinguish between variables and transients in this work. This is due to the slight observational ambiguity involved - e.g. the radio emission from a flaring star may indeed be seen in transient bursts, but the star itself is not truly a transient. The only unambiguous distinction between transients and variables is in the case of cataclysmic, one-off explosions (e.g. supernovae), but nevertheless our search methods would identify these, in addition to sources varying entirely above or around our detection thresholds.

\section{Feature Extraction}
\label{sec:features}

In order to make use of our anomaly detection tools, we must first calculate some features that describe our light curves. The choice of how to describe a dataset is crucial in machine learning tasks and can generally be split between features based on domain knowledge or experience and deep learning, where features are learned automatically from the data. 
In this work we use two domain-informed feature sets which have been used in machine learning searches for variability in previous studies. We also test the union of both these sets together, which we will refer to as the combined feature set. We choose to use these domain-informed features as opposed to those learnt from the data for several reasons. Firstly, automatic feature extraction methods can have many thousands of parameters (internal weights and biases) to update, requiring similarly extensive datasets. With our moderate sample of $\sim$9,000 light curves, taken from heterogeneously observed fields, we cannot expect our data to be representative of all possible underlying variability. Similarly, data-driven feature extraction can be computationally expensive, requiring many CPU- or GPU-hours to train. Finally, by using features tested in other regimes we can compare and contrast their performance between use cases. We will also compare the performance of the two feature sets to a baseline pair of parameters often used in transient detection and variable characterisation (see e.g. the ThunderKAT commensal work in section \ref{sec:intro}).

Every feature \textit{x} used in this work is standardised by computing

\begin{equation}
   z = \frac{x - \mu}{\sigma}
	\label{eq:normalise}
\end{equation}
i.e. by subtracting the mean $\mu$ and normalising by standard deviation $\sigma$. This is a common approach in machine learning applications (e.g. \citealp{Czech2018}) and is done so that features with large absolute ranges (or measured in different units) do not dominate. For example, if a model required computation of distances between datapoints in a 2-D dataset where one feature had a range $\mathcal{O}(10^4)$ whilst the other had a range $\mathcal{O}(1)$, the former would dominate the distance calculation.
\subsection{Baseline features}

As a baseline pair of features we make use of two parameters often used to discover radio transients. For a light curve  of $N$ flux density measurements $F_i \pm \sigma_i$, observed at frequency $\nu$, the two variability statistics are defined as

\begin{equation}
    \eta_\nu \equiv \chi^2_{N-1} =\frac{1}{N-1}\sum_{i=1}^N \frac{(F_{i,\nu} - \overline{F_\nu}^*)^2}{\sigma_i^2}
	\label{eq:eta}
\end{equation}

and

\begin{equation}
    V_\nu \equiv \frac{s_\nu}{\overline{F_\nu}} = \frac{1}{\overline{F_\nu}}\sqrt{\frac{N}{N-1}(\overline{F^2_\nu}-\overline{F_\nu}^2)}
	\label{eq:V}
\end{equation}
where $s$, $\overline{F_\nu}$ and $\overline{F_\nu}^*$ denote standard deviation, mean and weighted average of the flux density respectively. Broadly speaking, $\eta_\nu$, as a chi-squared test, quantifies the statistical significance of variability in a light curve,  whilst $V_\nu$ describes the amplitude of variability and is sometimes known as the modulation index. We note that this two parameter setup is far simpler (lower dimensions) than the two feature sets described subsequently, which may not capture more subtle variations in the light curves. This is due to our anomaly detection techniques being designed for high dimensional data and there is no guarantee that their algorithmic definitions of anomalies match human intuition when using $\eta_\nu$ and $V_\nu$. 

\subsection{Feets features}
\label{sec:feets}

Next we used a feature set from the \textit{feets} package \citep{Cabral2018ascl,Cabral2018}, inspired by the FATS code \citep{Nun2015}. This package combines over 40 features used for time series analysis, which can be broken into approximately three subsets. First, some features calculated are simple statistical quantities, such as the mean, amplitude, and standard deviation of a light curve. The second subset of features used are included due to their known statistical or astrophysical relevance to a given class of object such as variable stars \citep{Richards2011} and quasars \citep{Kim2011} and include, for example, the skewness and the exponent of a light curve's structure function. Finally, several features used are derived from a Lomb-Scargle periodogram \citep{Lomb1976, Scargle1982} and its resultant best fit to each light curve. We note that there are some features in this codebase that require colour information but as our data is univariate (monochromatic) we cannot employ these features in this implementation. These \textit{feets} features have an implementation built into \textsc{Astronomaly}'s framework, making them ready to test for a given application out-of-the-box. The full table of features used is detailed in Table \ref{table:feets}.

\label{table:feets}
\begin{table*}
    \centering
    \begin{tabular}{llll}
    \hline
    \textbf{Feature} & \textbf{Description} & \textbf{Inputs} & \textbf{Reference}  \\
    \hline
    \hline
    Amplitude & Half the difference between the median & $F_i$ & \cite{Richards2011}\\
    &  of the maximum 5$\%$ and the median \\
    & of the minimum 5$\%$ fluxes\\
    Anderson Darling test & A test statistic for non-Gaussianity &  $F_i$ & \cite{Kim2009}\\
     Auto correlation length & Length of linear dependence of a signal &  $F_i$ &  \cite{Kim2011} \\ 
     &with itself at two points in time, taken  \\
     &when the ACF reaches $1/e$\\
     Beyond1Std & Percentage of points beyond one &  $F_i$, $\sigma_i$ &  \cite{Richards2011}\\
     & standard deviation from the weighted mean \\
     
     Con & The number of three consecutive data & $F_i$ &   \cite{Kim2011} \\
     & points that are brighter or fainter than\\
     &2$\sigma$ and normalized by $N-2$ \\

     $\eta^e$ &$ \Bar{w}\; \sum w_{i}(F_{i+1} - F_i)^2 / (\sigma^2 \sum w_i)$& $F_i$, $t_i$ & \cite{Kim2014} \\
     &   where $w_i = 1/(t_{i+1}-t_i)^2$\\
     
     Fourier Amplitudes & $A_{i,j} = \sqrt{a_{i,j}^2 + b_{i,j}^2}$ & $F_i$, $t_i$ & \cite{Debosscher2007}\\
     & for $i = {1,2,3}$ and $j = {1,2,3,4}$\\
     
     Fourier Phases& $\phi_{i,j}=$arctan$ \left( \frac{b_{i,j}}{a_{i,j}} \right) - \phi_{1,1}$ & $F_i$, $t_i$ & \cite{Debosscher2007}\\
     & for $i = {1,2,3}$ and $j = {1,2,3,4}$\\
     
     Gskew & Median-of-magnitudes based & $F_i$ & \cite{Richards2011}\\
     & measure of the skewness \\
     Linear Trend & Slope of a linear fit to the light curve & $F_i$, $t_i$ &  \cite{Richards2011}\\
  
     Lomb-Scargle Period & The period of the largest & $F_i$, $t_i$ & \cite{Kim2011}\\
     & periodogram peak\\
     Lomb-Scargle $\eta^e$  & $\eta^e$  as applied to the phase folded light curve & $F_i$, $t_i$ & \cite{Kim2011}\\
     Lomb-Scargle $R_{CS}$ & $R_{CS}$ as applied to the phase folded light curve & $F_i$, $t_i$ & \cite{Kim2011}\\
     MaxSlope & Maximum absolute slope between & $F_i$, $t_i$ &  \cite{Richards2011}\\
     &two consecutive observations \\
     Mean $\Bar{F}$ & The mean flux - $1/N \; \Sigma_{i} F_i$ & $F_i$ &  \cite{Kim2014} \\
     Mean Variance & The ratio of the standard deviation &$F_i$ &  \cite{Kim2011}\\
& to the mean flux - $\sigma / \Bar{F} \equiv V$ \\
     Median Abs. Dev. & The median discrepancy of the data & $F_i$ &  \cite{Richards2011}\\
     & from the median - med$(|F_i - $med$(F_i)|)$ \\
     Median Buffer Range \% & Fraction of points with amplitude &$F_i$ &  \cite{Richards2011}\\
     & a tenth of the median flux \\
     Pair Slope Trend & The fraction of increasing first differences &$F_i$ & \cite{Richards2011} \\
     &  minus the fraction of decreasing\\
     &  first differences \\
     Percent Amplitude & Largest percentage difference between & $F_i$ & \cite{Richards2011}\\
     & either the max or min flux and the median. \\
     
     $Q_{3-1}$ & The difference between the 3rd & $F_i$ &  \cite{Kim2014} \\
     & and 1st quarterlies \\
      $R_{CS}$ & Range of cumulative sum &$F_i$ & \cite{Richards2011}\\
     Skewness & $\frac{N}{(N−1)(N−2)}\sum (F_i− \Bar{F} / \sigma)^3$ & $F_i$ & \cite{Richards2011}\\ 
     Slotted Auto Correlation  & Slotted auto correlation length & $F_i$, $t_i$ & \cite{Protopapas2015}\\ 
     Function Length \\
     Small Kurtosis & $E[(F-\Bar{F} \:/\sigma)^4]$ & $F_i$ & \cite{Richards2011}\\ 

     Standard Deviation & Standard deviation of fluxes &$F_i$ & \cite{Richards2011}\\ 
     Structure Function &  The exponent of the $SF \propto \tau^{\beta}$  & $F_i$, $t_i$ & \cite{Hughes1992}\\
     Index & where $SF(\tau)=\langle[F(t) −F(t+\tau)]^2\rangle$ \\
    \hline
    \end{tabular}
    \caption{Features used in this work as calculated by the \textit{feets} package \citep{Cabral2018ascl}. Each light curve consists of $N = \Sigma i$ measurements of flux density $F_i \pm \sigma_i$ taken at time $t_i$. A similar version, for the original FATS code is described by \citet{Webb2020}.}
    \label{tab:features}
\end{table*}

\subsection{Wavelet features}
\label{sec:wavelets}

The final set of features used come from studies primarily aimed at describing and classifying supernovae \citep{Lochner2016, Alves2022}, though they have also been used in the classification of radio transients \citep{Sooknunan2021}. These features, called wavelet features, arise from the understanding that time series $f(t)$ can be decomposed into a sum of basis functions

\begin{equation}
    f(t) = \sum_i a_i \phi_i(t)
	\label{eq:phi}
\end{equation}
with coefficients $a_i$, where $\phi_i(t)$ might for example be a set of sines and cosines (i.e. a Fourier decomposition). We can then use the coefficients of our basis functions as a set of features for later use. Here, we use the symlet family of wavelets as our basis functions (see e.g. \citealp{arfaoui2021wavelet} for more on these),  due to their successful use in previous work \citep{Lochner2016, Alves2022}. These wavelet features are calculated with minimal assumptions on the data and so it is our hope that they describe the observed light curves well,  whilst by contrast the feets package's parametric calculations are made with assumptions on the statistical distributions of the data. Furthermore, wavelet transforms are translation equivariant and so the same object (e.g. a SN) observed at different times since peak brightness should produced similar coefficients. 

We calculate the wavelet coefficients for our light curves using the \textsc{snmachine} codebase developed by the LSST Dark Energy Science Collaboration \citep{Lochner2016, Alves2022}. 
In order to do this, we first interpolate our light curves with a Gaussian Process Regression \citep[GP;][]{Rasmussen2006}, allowing us to account for gaps in our uneven time series. To do this, all light curves are shifted to take place between $0$ and $t_{max}$ days, where for our dataset $t_{max}$ = 1300 days, corresponding to our longest light curves in the GX339$-$4 field.
The underlying procedure built into \textsc{snmachine} fits Gaussian processes from the \texttt{George} library\footnote{\url{https://george.readthedocs.io/en/latest/}}, with this implementation using GPs with a mean of zero and one of two covariance kernels. The two kernels tested are an exponential square kernel of the form 
\begin{equation}
     k_{\mathrm{exp}}(t_i,t_j) = \mathrm{exp}\left(-\frac{|t_i - t_j|^2}{2}\right),
\end{equation}
and a combination of $k_{\mathrm{exp}}(t_i,t_j)$ with an exponential sine squared kernel of scale parameter $\Gamma$ and period $P$ i.e.
\begin{equation}
    k_{\mathrm{sin}}(t_i,t_j) = k_{\mathrm{exp}}(t_i,t_j) + \mathrm{exp}\left(-\Gamma \sin^2\left[\frac{\pi}{P}|t_i - t_j|\right] \right).
\end{equation}
These two kernels are intended to capture any periodic behaviour whilst the exponential decays account for the idea that points separated further in time are expected to be less correlated. The kernel used is determined by minimising the negative log-likelihood of the GP with the SciPy implementation of a limited-memory Broyden–Fletcher–Goldfarb–Shanno algorithm \citep{Byrd1995SJSC...16.1190B, 2020SciPy-NMeth}. Once the Gaussian Process has been fit, we evaluate the regression every $\sim$5 days (i.e. at 276 instances), approximately matching our typical observational cadence. Example Gaussian process regression fits can be seen in Figure \ref{fig:light_curves}. For light curves shorter than
$t_{max}$ in length, the remaining datapoints are assumed to be 0.

Once all light curves are interpolated by the regression, the wavelet transform is calculated, producing a set of 275 highly redundant coefficients. In order to reduce this high dimensionality, we run a Principal Component Analysis \citep[PCA;][]{Pearson1901}, which transforms the data into an orthogonal basis. In this now-orthogonal data space, we choose to preserve the 40 largest eigenvalues for each light curve, which captures 99.999995 per cent of the summative variance of the data. This results in reducing the data space from 276 features per light curve to only 40, while retaining almost all of the variance. 
Fewer principal components would result in faster downstream computations, however as mentioned earlier, anomalies (transients) constitute a small minority of our sample whilst the majority are stable light curves which should require few wavelet features to describe. Furthermore, light curves shorter than $t_{max}$ are assumed to be 0 past their final datapoint, which will contribute towards the majority of the dataset being non-variable. As a result, initial testing with only 10 principal components showed markedly worse performance. This shows that the diversity of light curve lengths affects the resultant wavelets and their principal components retained, which may bias results.
The choice of 40 principal components is a balance between reducing the dimensionality of the features whilst trying to preserving enough information such that the anomalies, which constitute a small minority of all of the light curves, are retained. 
In a truly unsupervised setting, it is not possible to know how many eigenvalues to preserve as one has no labels to train from.

Finally, it is worth noting that extracting this set of features is the most computationally expensive part of our analysis. However, even the slowest step, the Gaussian process regression, runs in 10 minutes for the entire database on a standard desktop using a 3.5~GHz quad-core processor. Also, all of these features, once calculated, can be saved and these routines then need not be run again, making static comparisons like this work fast to conduct. For a real-time system features may be re-calculated for every new datapoint, however this may not be optimal or required. For example, the Gaussian Process Regressions that underlie the wavelet features scale naïvely as $\mathcal{O}(N^3)$ and so calculating them for each new datapoint may be prohibitive. For larger datasets or for a real-time transient detection system a balance between computational scalability and information gained may be required and could dictate what feature sets can be deployed for a given science case.

\subsection{Feature space and clustering}
\label{sec:features-results}

An obvious question to ask is how well the feature sets and baseline feature pair separate transients (anomalies) from the rest of the distribution. For a perfect feature set, we would want all transients to be separated cleanly from all non-anomalies. To inspect this we make use of a Uniform Manifold Approximation and Projection (UMAP, \citealp{McInnes2018}), a technique for dimensionality reduction. Briefly, the UMAP algorithm works by building a topological representation of data in its N-dimensional space (where \textit{N} is 45 and 40 for the \textit{feets} and wavelet features respectively). Once this has been constructed, the UMAP algorithm  works to optimize a low dimensional (e.g. 2-D) representation of this topological space as measured by cross entropy. Datapoints that are similar in the N-D space should then remain nearby in the 2-D representation, which can then be used for visualisation or further downstream tasks.

\begin{figure*}
    \includegraphics[width=\textwidth]{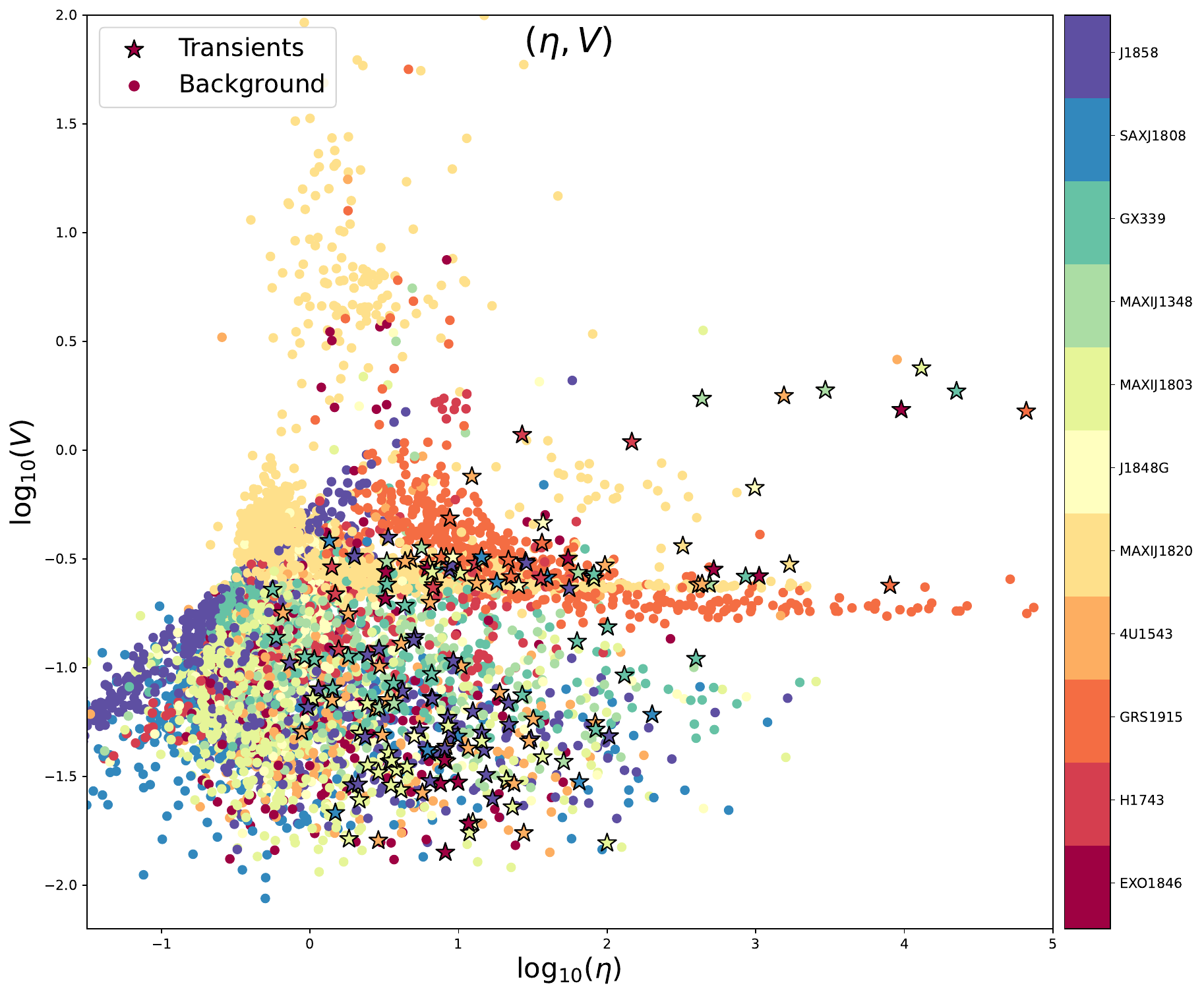}
    \caption{The ($\eta, V$) feature space, colour-coded by observational field. Stars indicate volunteer-verified anomalies (transients). There are clear field-dependent systematics affecting, for example, the MAXI~J1820+070 field.}
    \label{fig:EtaV}
\end{figure*}

As a baseline, the feature space of ($\eta, V$) is already in 2-D so there is no need to reduce its dimensionality. This feature space is shown in Figure~\ref{fig:EtaV}. This figure demonstrates that many transients and variables are inliers, as found in \citetalias{Andersson2023}. The presence of systematics in some fields (e.g. MAXI~J1820) can be seen, along with the difficulty in separating our transients (anomalies) from all background sources. This was one of the main motivations for testing the subsequent feature sets for anomaly detection.

\begin{figure*}
	\includegraphics[width=\textwidth]
    {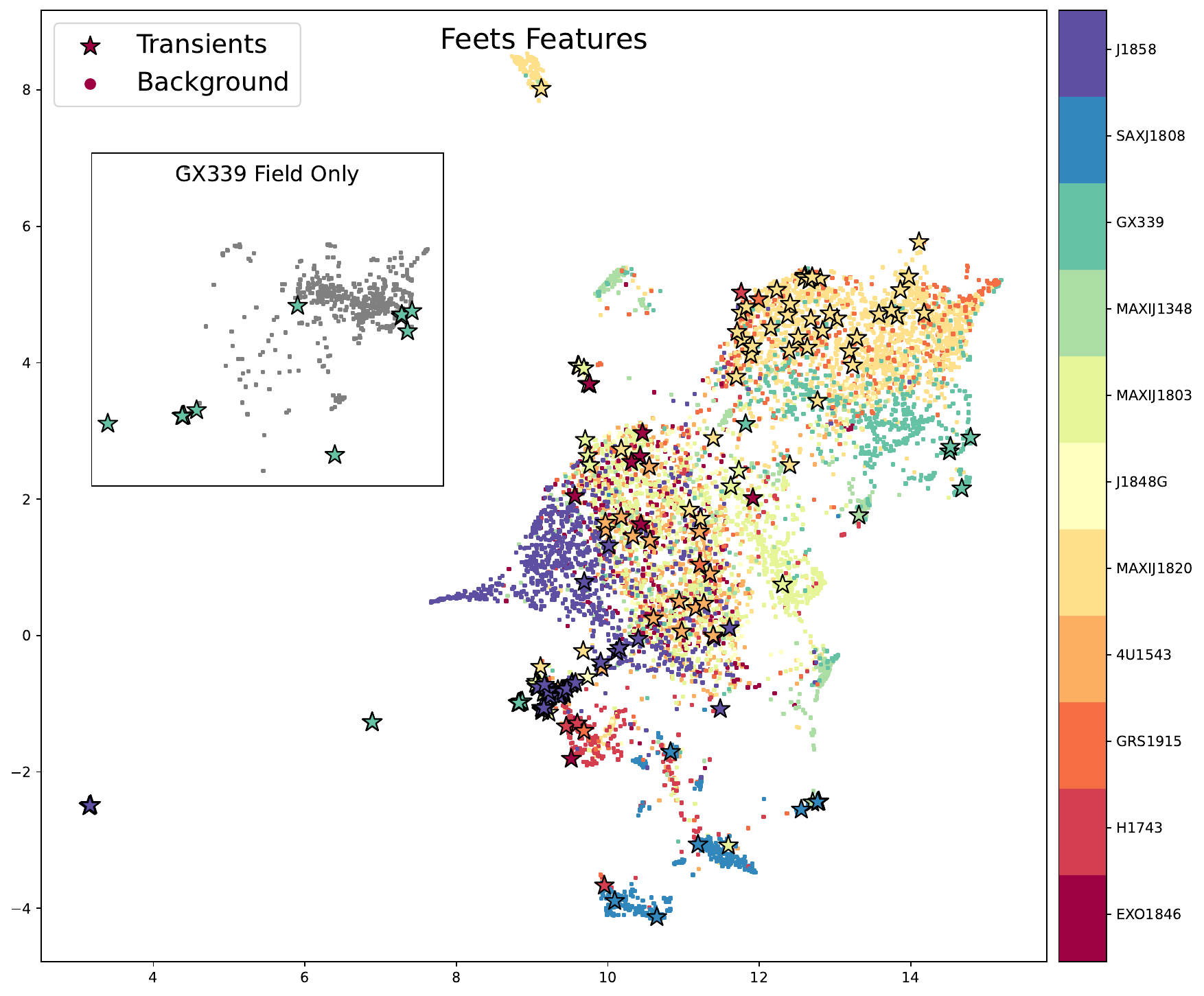}
    \caption{The UMAP of our \textit{feets} feature space, colour-coded by observational field. We see that clustering is mainly a function of observational field, aligning with the heterogeneous nature of the dataset. The axes are unlabelled as these numbers are arbitrary combinations of many features and have no physical meaning. The inset is a demonstration of how the data of one particular field (GX339) is distributed, as mentioned in the text.}
    \label{fig:UMAP-feets}
\end{figure*}

The \textit{feets} feature space for our dataset can be seen in Figure \ref{fig:UMAP-feets}, where stars are our citizen-scientist verified anomalies (transients) and the colourbar indicates which observational field a given light curve is from (see section \ref{sec:obs}). We can see that sources in a given field are clustered together. This is likely due to observational systematics varying on a field-by-field basis. For example, the SAX~J1808 field only received 6 observations and, resultingly, exists in a very different area of parameter space to other fields. Whilst this might be of slight concern, we do not use these 2-D representations for downstream tasks, but rather as sanity checks and in order to judge qualitatively how well our data (anomalies/transients and everything else) are described by a given feature set. Moreover, for a real-time or low-latency detection system, what is of more importance is how well transients separate from background data \textit{within} a given field - as in such a system, it is only against sources in the same observation that we would compare. Indeed, we can see that anomalies are somewhat clustered and that, whilst they tend to lie in their respective field of light curves, they are often on the edges of each region. A demonstration of this can be seen for the GX339$-$4 field shown in the inset of Figure \ref{fig:UMAP-feets}.

\begin{figure*}
    \includegraphics[width=\textwidth]{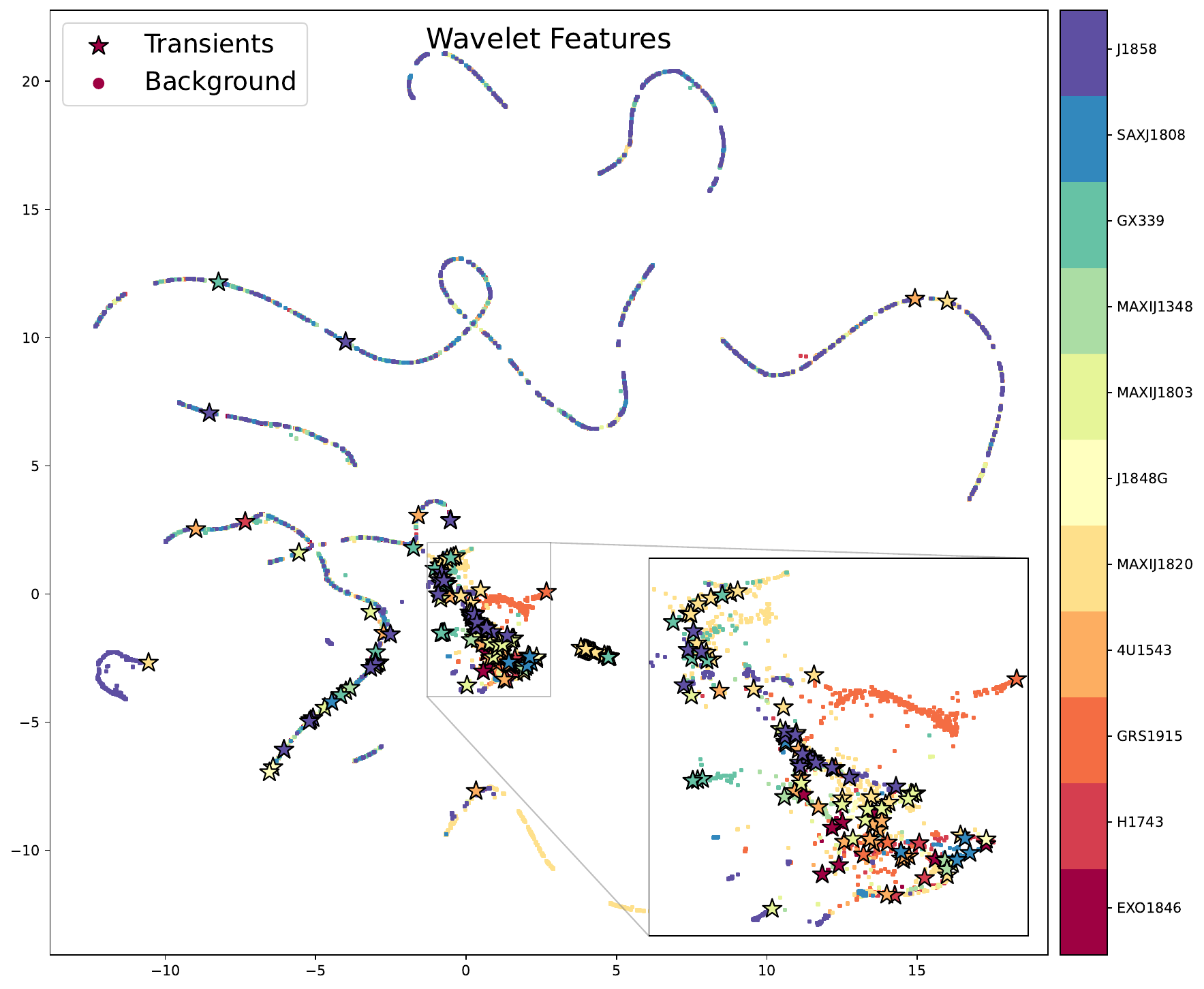}
    \caption{The UMAP of our wavelet coefficient feature space, colour-coded and labelled as in Figure \ref{fig:UMAP-feets}. In this instance, the observational fields with longer light curves (e.g. J1858, GX339$-$4) show distinct patterns, whilst most other sources are in a very dense region of parameter space, as shown by the inset zoom.}
    \label{fig:UMAP-WV}
\end{figure*}

Similarly, the feature space of our wavelet coefficients in Figure~\ref{fig:UMAP-WV} shows field-dependent clustering. However, in this case, there are clear strings of points in feature space that correspond to the observational fields with longer light curves (e.g. J1858, GX339$-$4). The majority of these sources are not transient and so perhaps these strings of points arise from requiring the same number of coefficients, of similar numerical values, to reproduce similar light curves. By contrast, shorter light curves and variable sources are clustered in the same area of parameter space (inset), however as are some non-anomalous background sources. This distinct cluster of data points may give insight into the performance of the wavelet features in \textsc{Astronomaly}'s models in sections \ref{sec:anomaly} and \ref{sec:active}.

\section{Anomaly detection and active learning}
\label{sec:anomaly}

Having calculated features for our light curves, we move on to the core anomaly detection step of \textsc{Astronomaly}. To do this we pass our standardised features into two commonly used anomaly detection algorithms, detailed below, as implemented in the \texttt{scikit-learn} package \citep{Pedregosa2011}. Anomaly detection is typically done by each algorithm having a definition of what is normal and assigning subjects that are far in parameter space from normal as highly anomalous, without any user labels or training. Once the degree to which something is anomalous has been ranked by a given algorithm, \textsc{Astronomaly} scales this to a human reading of 0 to 5, where 5 is most anomalous.
The output algorithm scores are linearly rescaled such that 5 represents the most anomalous source while 0 is the least.

By its nature, unsupervised learning has no pre-defined metric to calculate success. However, as noted earlier, we have votes from citizen scientists for all of our light curves from BfS:MKT, which in turn have been verified by us. Throughout this work we will assume that these verified volunteer votes are an absolute ground truth against which to compare our anomaly detection methods. We can then use the positions of our citizen-science verified transients in our anomaly score-ordered list to calculate metrics of success - the more transients that are ranked as highly anomalous, the better. 
It is important to point out that \textsc{Astronomaly} never decides whether or not something is anomalous and simply orders the entire list by anomaly score. The use of information from our volunteers is what allows us to then calculate metrics of success such as recall and precision (see section \ref{sec:recall_precision}). A contrasting method for cases without such citizen science data might be to define some threshold in an anomaly score-ordered list above which things are said to be anomalies, providing the binary classes on which to calculate performance.

\subsection{Anomaly detection algorithms}

The first of the two anomaly detection methods used herein is the Local Outlier Factor (LOF, \citealp{Breuniq2000}). The LOF is a density-based method that measures the deviation of the density of a given sample with respect to some user-defined number of nearest neighbours, $k$. We set $k = 100$ throughout this work, approximately 1 per cent of our sample. Values of $k$ between 20 (the default) and 200 were tested, producing changes in performance at the sub per cent level. As noted earlier, without labelled data it would be difficult to optimise this (and other) hyperparameters. The LOF is local in that the resultant anomaly score depends on how isolated the object is with respect to the surrounding neighbourhood. More specifically, the LOF of datum A is a measure of the average local density of all \textit{k} neighbours divided by A's own local density. One advantage of this local approach is that it might pick up anomalies missed by global density measurements. That is, a point only a small distance from a very dense cluster might be an anomaly, whereas a point in a sparse cluster might still have the same density as its neighbours and so not be an outlier despite its low absolute density value. For more on the distinction between global and local anomalies, see the discussion in \cite{Rogers2024}.

The second method used is Isolation Forest (IF, \citealp{Liu2008}). IFs take inspiration from the supervised random forest algorithm \citep[RF,][]{Breiman2001} and operate on the core principle that anomalies are both few in number and different compared to normal data. In an IF an ensemble of decision trees randomly split random features, growing each ‘branch' in the tree until all considered data are isolated. Given that anomalies should be few and different, they are easier to isolate than normal data that are close to each other in feature space. Therefore data with shorter decision branches are more likely to be anomalous, whilst those that take many splits to isolate are more similar to the rest of the sample. The final anomaly score is an average score over all trees in the forest. This model is particularly applicable to high dimensional data as there is no costly calculation of distances and therefore has a linear time complexity (i.e. twice the input data takes twice the runtime), c.f. LOF which can take up to $\mathcal{O}(N^2)$ \citep{Breuniq2000}.

\subsection{Anomaly Recall and Precision}
\label{sec:recall_precision}

 In order to assess the performance of our models we can ask out of all the anomalies that exist, what fraction did we find in the top \textit{n} sources? This quantity is known as the recall (also referred to as the completeness, true positive rate or the sensitivity) and is only calculable with ground truth labels for an entire dataset, as without this it is impossible to know how many anomalies are missed (false negatives). Recall is defined as 
 \begin{equation}
         R = \frac{\mathrm{True~Positives}}{\mathrm{True~Positives~+~False~Negatives}},
    \label{eq:recall}
 \end{equation}
 where for us the denominator is our total number of transients (anomalies) in the data, i.e. 168. This recall value, whilst simple, corresponds well with preferable performance consisting of many more anomalies high in the list for the user to see quickly. We can calculate the recall $R$ as a function of \textit{n} until we reach the end of our dataset ($n = 8874)$, by which time we will have recovered all sources in the list. We define $R_{10}$ and $R_{30}$ as the recall of transients in the top 10 and 30 per cent of our list respectively, which we will use as quantitative measures of performance throughout this work. We stress again that it is impossible to calculate this recall without labels for all transients, as without this we would not know how many transients fall below a given threshold i.e. the rate of false negatives. As such, looking at the top 10 or 30 per cent of the anomaly score-ordered lists is somewhat arbitrary and depends on, for example, the resources available for follow up or further citizen science classification.

We can also ask what fraction of the top $n$ sources are transients. This is known as the precision, also referred to in astronomical literature as the purity, defined as 
 \begin{equation}
         P = \frac{\mathrm{True~Positives}}{\mathrm{True~Positives~+~False~Positives}}.
    \label{eq:precision}
 \end{equation}
This can be thought of as a measure of the hit-rate of our anomalies. Namely, how many transients (anomalies) would there be in the top 10 or 30 per cent of the data?  By definition, as $n$ approaches the total length of our dataset (8874 sources), the precision will tend towards the aggregate rate of transients in our data i.e. $\sim2$~per~cent. precision, unlike recall, only depends on the labels of sources above some cut-off point set by the user and so is calculable even for a largely unlabelled dataset.

\subsection{Active Learning} 
\label{sec:active}
\begin{figure*}
	\includegraphics[width=\textwidth]{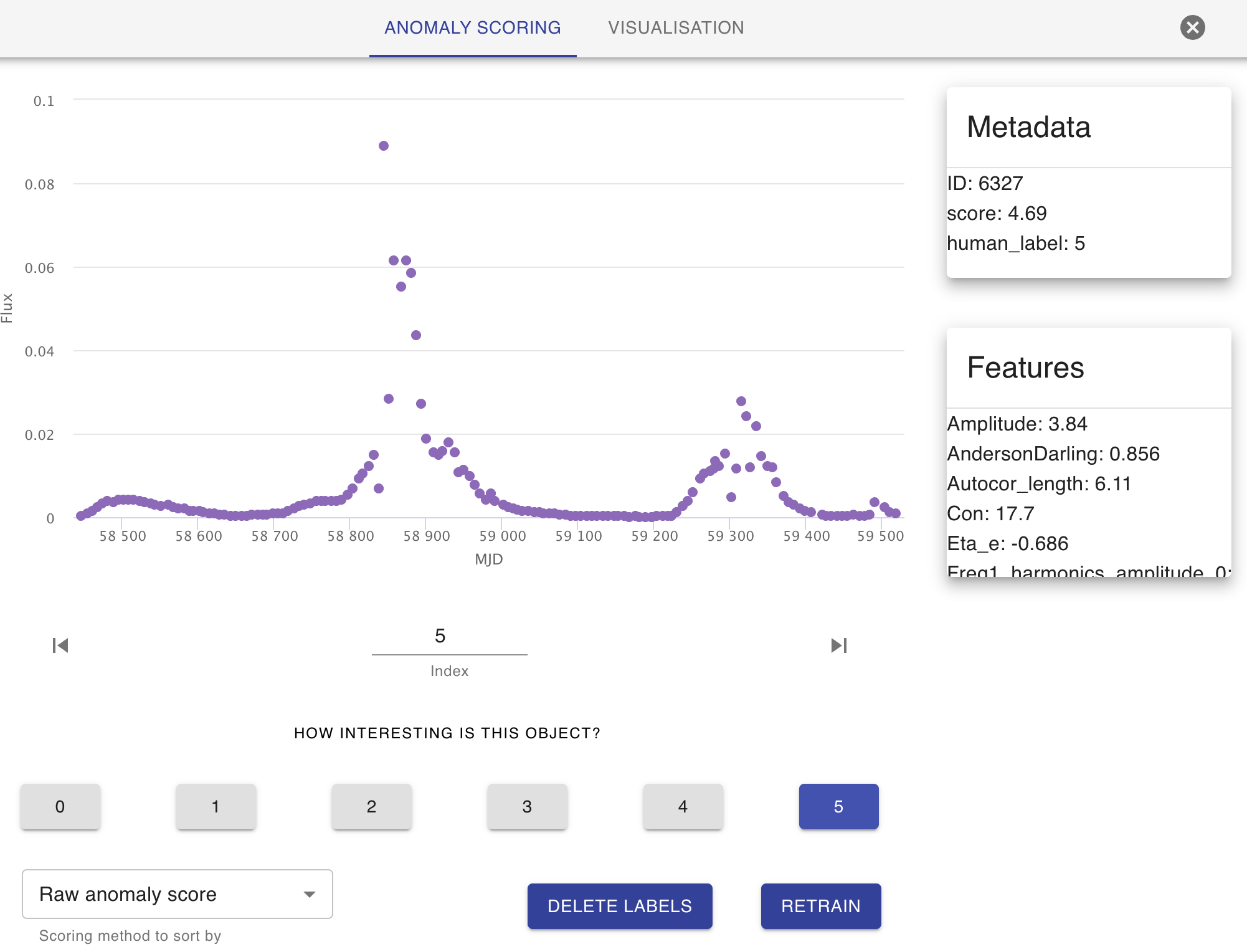}
    \caption{The \textsc{Astronomaly} GUI, showing subjects in anomaly-order. Using this interface, we can easily inspect our data and provide labels on some subset of objects in order to perform active learning.}
    \label{fig:GUI}
\end{figure*}

Once anomaly scores have been calculated for all data, \textsc{Astronomaly} allows for interactive visualisation by use of a GUI, seen in Figure \ref{fig:GUI}. This consists of sources listed ordered by anomaly score, displaying the raw data, its features and any associated metadata. From this interface we can label sources on a range of 0 to 5. This allows for active learning, whereby an additional model balances the raw anomaly score with user input and calculates a trained anomaly score. This then acts as a recommendation engine, displaying sources with higher trained anomaly scores preferentially to the user, speeding up anomaly detection.

The core of this active learning process is described by ranking objects with a new score

\begin{equation}
    \hat{S} = S   \mathrm{tanh}(\delta - 1 + \mathrm{arctanh}(\Tilde{U}))
    \label{eq:active}
\end{equation} 
where $S$ is the default anomaly score, $\delta$ represents a distance penalty term and $\Tilde{U}$ describes a relevance score given by 

\begin{equation}
    \Tilde{U} = \epsilon_1 + \epsilon_2\left(\frac{U}{U_{\mathrm{max}}}\right).
    \label{eq:relevance}
\end{equation}
Here $\epsilon$ are normalisation constants and the user input score is $U$, normalised by $U_{\mathrm{max}} = 5$. Since $U$ will only ever be input by the user for a small fraction of a dataset, its value for all remaining sources is estimated by a random forest regression calculated with 100 estimators. This regression will have uncertainty, quantified by $\delta$, which is large when an object is far from a human-labelled neighbour and small when near to labelled data. In summary this means that, in regions far from human-labelled data ($\delta \gg 1$), $\hat{S}$ reverts to its original score as tanh~$\rightarrow 1$.
Similarly, if $U\sim U_{max}$ then $S$ will not change due to high human ranking. Conversely, regions of feature space near labelled data ($\delta \ll 1$), with low user scores will be downweighted, corresponding to relegating uninteresting data to lower in our anomaly-ranked list.

In this work we actively label the top $\sim$ 2 per cent of each feature-model pair, which takes of order 10 minutes by an individual (AA), applying a consistent methodology throughout. That is, examples of clear variability or known transients were ranked highly ($U = 4$ or 5), whilst low-significance light curves, or those showing clear false positive behaviour - e.g. where one epoch shows large changes in flux due to a different calibrator or worse image quality - were ranked at 1 or 0. 

The rate of retraining in \textsc{Astronomaly} is entirely determined by the user(s). Any number of sources can be labelled between training iterations, with between 10 and 20 sources proving to be a qualitatively good balance between providing enough data for training whilst updating regularly enough for efficient labelling. As an example of this procedure, if the user labelled a few sources that were deemed uninteresting in the same way this would, upon retraining, down-weight further instances of similar light curves (that is, ones with similar features). However, it is worth noting that this labelling is inherently subjective and dependent on a user's end goal. An instrument scientist interested in flux calibration would score sources differently to an astronomer interested in stellar radio flares. Active learning attempts to quantify this and balance an algorithmic definition of anomalous with a user-dependent idea about what is an \textit{interesting} anomaly. Finally we note that the regressor used to model user interest is by default a RF but that using a Gaussian process provides an improvement on this for some datasets, which has been incorporated into the most recent versions of \textsc{Astronomaly} \citep{Walmsley2022b, Lochner2025}.


\section{RESULTS}
\subsection{Baseline Anomaly Detection}
\label{sec:anomaly-results}

\begin{figure*}
	\includegraphics[width=0.24\textwidth]{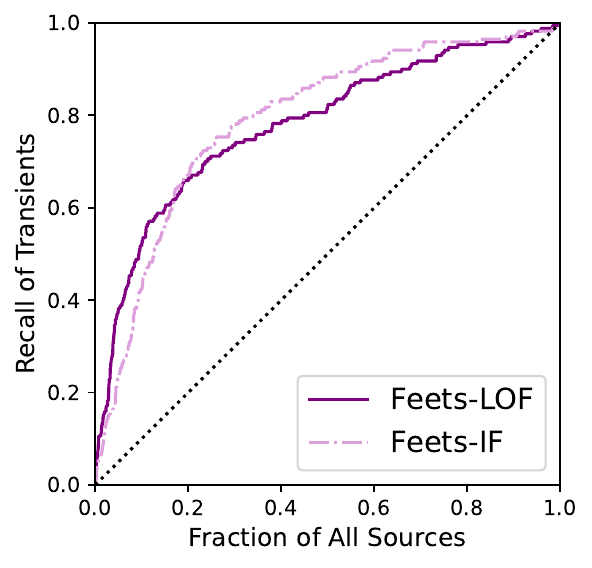}
 	\includegraphics[width=0.24\textwidth]{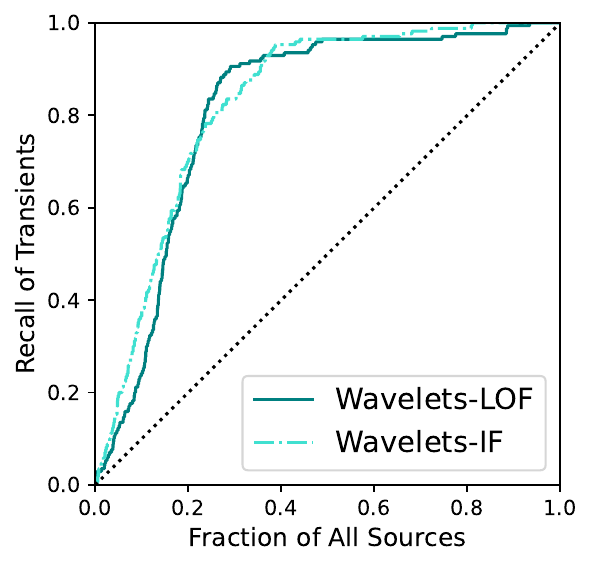}
     	\includegraphics[width=0.24\textwidth]{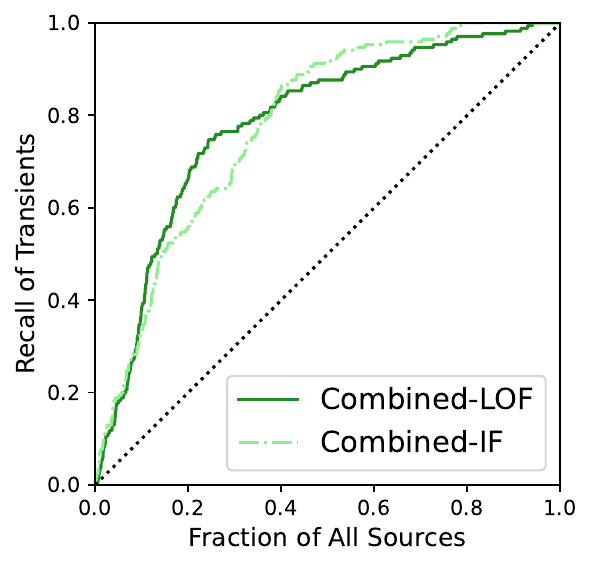}
	\includegraphics[width=0.24\textwidth]{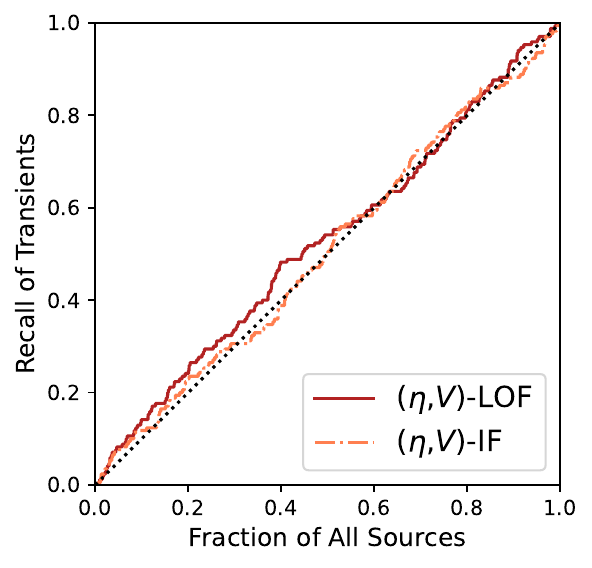}
 	\includegraphics[width=0.95\textwidth]{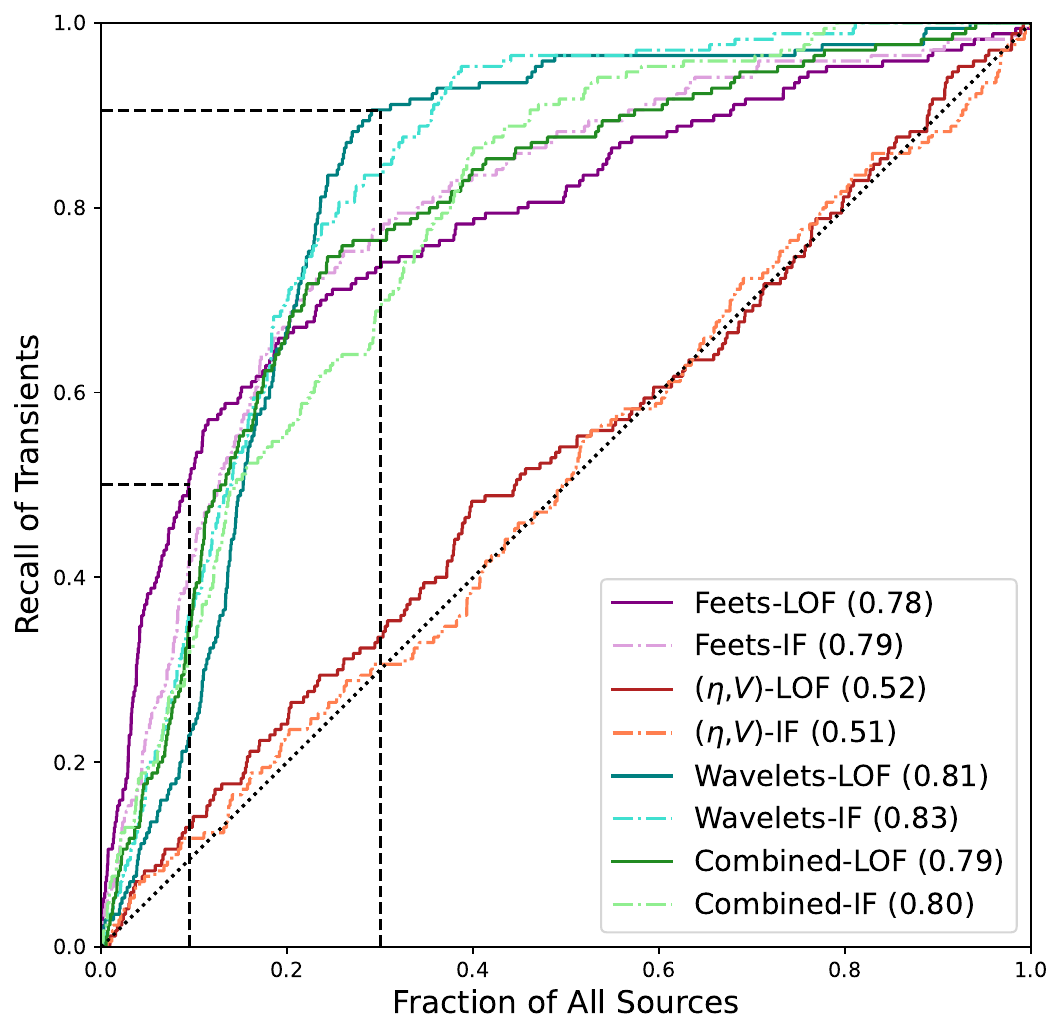}
    \caption{\textbf{Upper panel:} The recall curves for both anomaly detection algorithms as applied to four groups of features derived from our radio light curves. The data are ordered according to the anomaly score from each algorithm, where for example $x=0.3$ corresponds to the most anomalous thirty per cent of the dataset. The dotted line in each corresponds to $y=x$ i.e. random chance.
    \textbf{Lower panel}: A combined plot of all eight recall lines for better comparison between feature sets. Dashed lines make comparisons (see text) for recall rates at 10 and 30 per cent of the dataset. The values in the legend correspond to the area under each recall curve (AUC).}
    \label{fig:recalls}
\end{figure*}

The recall curves for both anomaly detection algorithms applied to all four feature sets can be seen in Figure \ref{fig:recalls}. These are normalised on both axes i.e. x=0.1 corresponds to the top 10 per cent most anomalous sources and y=0.5 corresponds to half of the bona fide 168 variables found by volunteers. Better performance corresponds to curves with higher gradient at low $n$, appearing pushed up into the upper left corner. It can be clearly seen that ($\eta, V$) show poor performance on this dataset, essentially matching random chance, in line with \citetalias{Andersson2023}'s findings. As noted earlier, this will be due to a combination of applying a different methodology to the intended use of $\eta$ and $V$ and the inherent biases of these two statistics - e.g. $\eta$ is correlated with signal-to-noise squared (see equation \ref{eq:eta}). Furthermore, our ground truth data consist of sources chosen irrespective of their ($\eta, V$) values. Contrastingly, we see that our main feature sets are able to recover a majority of our volunteer-verified transients (anomalies) in a minority of the data. This is the first demonstration of anomaly detection as applied to radio time series.

When comparing performance between anomaly detection models (i.e. for the same feature set), we see minimal variation and qualitatively similar behaviour between recall curves. For the \textit{feets} and wavelet features we see ‘broken power-law' type behaviour, with an approximately constant gradient initially, followed by a knee and then worse performance. When combining these feature sets, performance is similar, though the IF and LOF diverge more after $x=0.1$ before reconverging. Table \ref{tab:recalls} lists all $R_{10}$ and $R_{30}$ values for feature-model set pairs. The largest discrepancies between models occur at the ‘top' of our list, with $R_{10}$ values varying by 0.1 for \textit{feets} features and 0.14 for wavelets. By contrast, $R_{30}$ values differ from 0.03 to 0.07, indicating that model choice most affects the highly anomalous sources in our dataset. When combining the \textit{feets} and wavelet features, this difference between $R_{10}$ and $R_{30}$ values is less pronounced. It is worth noting that several iterations of each feature-model pair were tested, with performance never varying by more than $\sim$1 per cent. As a result, all values are rounded to this precision.

Differences in feature set have a much greater effect than changing the anomaly detection method - for the same model $R_{10}$ changes by factors of $\sim$3.5 between best- and worst- performing feature sets. Similarly, $R_{30}$ for the baseline parameters reaches only $\sim$0.34, whilst the wavelet features can recover up to 91 per cent of volunteer-verified transients. This underscores the importance of using features that best describe a given dataset.

We can also calculate an aggregate recall performance by simply integrating the area under the curve (AUC) of each feature-model pair's recall curve. The corresponding AUC values for each feature-model instance are listed in the legend of Figure \ref{fig:recalls}. As already mentioned, differences between feature sets are greater than differences between anomaly detection models. These AUC values show that the wavelet-IF combination has the best recall performance at an AUC of 0.83. However, this is a summary over all the data and does not consider the actual finding of transients in a live setting. By this we mean that one must consider how many anomalies a human will be able to find in a reasonable amount of time. As discussed above and can be seen in the lower panel of Figure \ref{fig:recalls}, if a user can only inspect the top 10 per cent of the dataset, then the feets-LOF analysis performs the best, with $R_{10} = 0.52$. However, if there are resources available to inspect 30 per cent of the dataset (e.g. computation, citizen scientists, a team of experts) then the LOF performance on wavelet features recalls 91 per cent of transients, 7 per cent better than the next-best feature-model pair ($R_{30} = 0.84)$.

\begin{table*}
\centering
\begin{tabular}{|c|cc|cc|cc|cc|}
\hline
 & \multicolumn{2}{|c|}{\textit{feets}}    & \multicolumn{2}{|c|}{Wavelets} & \multicolumn{2}{|c|}{$(\eta, V)$} & \multicolumn{2}{|c|}{Combined}   \\ 
 & \multicolumn{1}{|c|}{LOF} & IF & \multicolumn{1}{|c|}{LOF} & IF & \multicolumn{1}{|c|}{LOF} & IF & \multicolumn{1}{|c|}{LOF} & IF \\ \hline
$R_{10}$ & \multicolumn{1}{|c|}{0.52, \textbf{0.52}} & 0.42, \textbf{0.45} & \multicolumn{1}{|c|}{0.24, \textbf{0.32}} & 0.38, \textbf{0.42} & \multicolumn{1}{|c|}{0.14, \textbf{0.16}} & 0.12, \textbf{0.11}     & \multicolumn{1}{|c|}{0.38, \textbf{ 0.51}} & 0.34, \textbf{0.37}                 \\ \hline
$R_{30}$ & \multicolumn{1}{|c|}{0.74, \textbf{0.73}} & 0.78, \textbf{0.78} & \multicolumn{1}{|c|}{0.91, \textbf{0.91}} & 0.84, \textbf{0.84} & \multicolumn{1}{|c|}{0.34, \textbf{0.32}} & \multicolumn{1}{|c|}{0.31, \textbf{0.30}}  & \multicolumn{1}{|c|}{0.76, \textbf{0.78}} & 0.69, \textbf{0.74} \\ \hline \hline

$P_{10}$ & \multicolumn{1}{|c|}{0.11, \textbf{0.11}} & 0.09, \textbf{0.10} & \multicolumn{1}{|c|}{0.05, \textbf{0.06}} & 0.07, \textbf{0.08} & \multicolumn{1}{|c|}{0.03, \textbf{0.03}} &0.02, \textbf{0.02}   & \multicolumn{1}{|c|}{0.07, \textbf{0.10}} & 0.07, \textbf{0.07}                    \\ \hline
$P_{30}$ & \multicolumn{1}{|c|}{0.05, \textbf{0.05}} & 0.05, \textbf{0.05} & \multicolumn{1}{|c|}{0.06, \textbf{0.06}} & 0.05, \textbf{0.05} & \multicolumn{1}{|c|}{0.02, \textbf{0.02}} & \multicolumn{1}{|c|}{0.02, \textbf{0.02}} & \multicolumn{1}{|c|}{0.05, \textbf{0.05}} & 0.04, \textbf{0.05}  \\ 
\hline
\end{tabular}
\caption{Specific recall ($R$) and precision ($P$) values, rounded to two decimal places, for each feature-model pair, measured after ten and thirty per cent of the most anomalous sources have been ‘seen'. Each pair of values listed are the default, \textbf{active} iteration of that feature-model instance.}
\label{tab:recalls}
\end{table*}

The precision values for all feature-model pairs can be found in Table \ref{tab:recalls}, again looking at 10 and 30 per cent of our anomaly score-ordered lists. For the \textit{feets} and wavelet feature sets, precision is always at least a factor of 2 better than the corresponding baseline performance. In general, the same trends hold as with our recall values - differences between feature sets are greater than differences between anomaly detection models. Likewise, we can see that the \textit{feets} features produce the best results when looking at the top 10 per cent of the data, with a hit-rate of over 1 in 10, compared to the baseline of 1 in 50. The results 30 per cent of the way through the anomaly score-ordered lists, as seen with the recall values, show that the wavelet-LOF instance performs the best, but only marginally. Nevertheless, we can see that for either feature set or their combination, as applied to both of our anomaly detection algorithms, the hit-rate of anomalies in the data is always a factor of 2 or greater when compared to our baseline. This directly translates into the number of sources that would need to be sifted out before finding a new transient.

In summary, the choice of model used makes a small difference, though LOF is marginally preferred. When considering feature sets, the baseline features perform the worst, recovering transients almost uniformly through the anomaly score-ordered lists. The \textit{feets} features perform the best at the beginning of each anomaly score-ordered list, recovering 52 per cent of transients in 10 per cent of the data and increasing the precision above random by a factor of 5. However, the wavelet features perform better when considering a larger fraction of each anomaly score-ordered list. Using the wavelet features over 90 per cent of anomalies are recovered in 30 per cent of the data, at a precision level that is three times higher than the baseline features.

\subsection{Active Learning Improvement}
\label{sec:active-results}

\begin{figure*}
	\includegraphics[width=\textwidth]{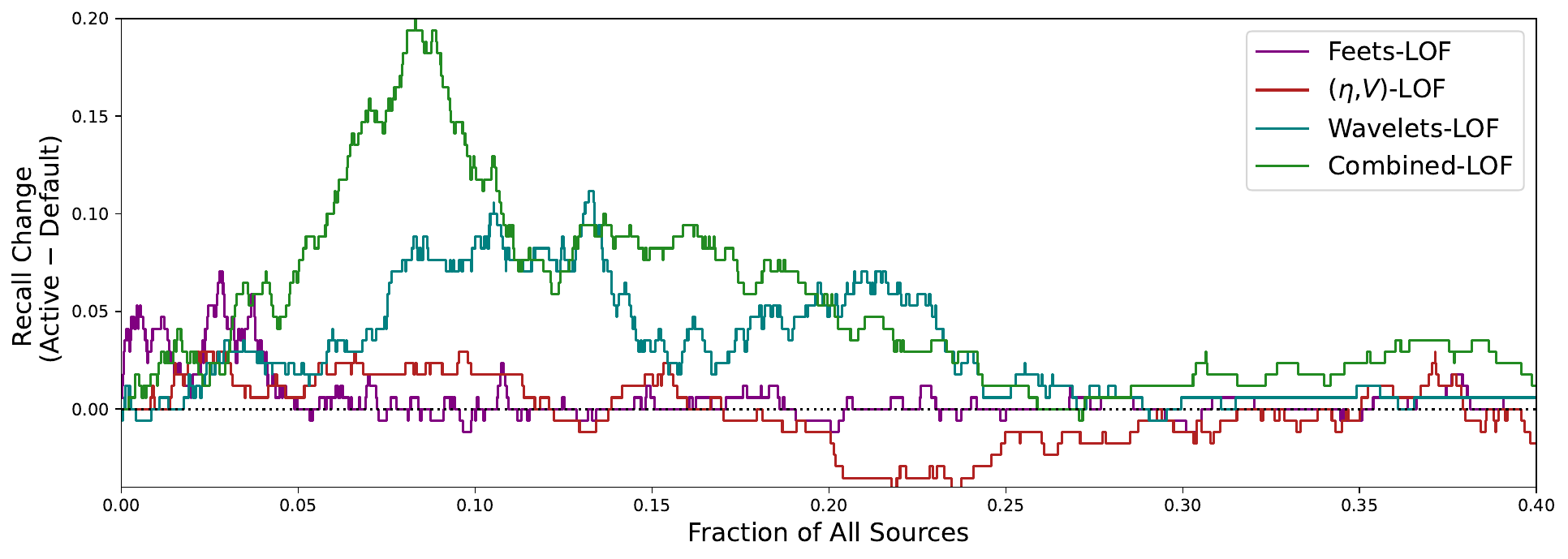}
 	\includegraphics[width=\textwidth]{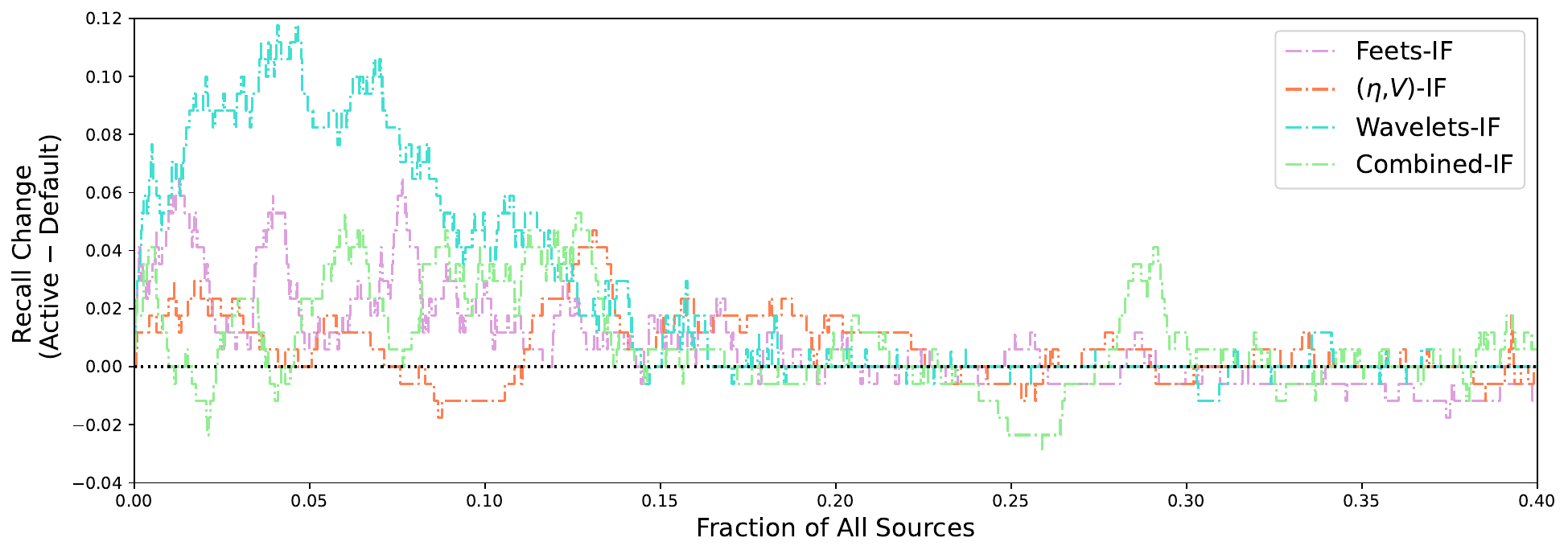}
    \caption{Changes in performance due to active learning for each feature-model pair, where $y=0$ indicates no improvement. The x-axis refers to how far down each anomaly score-ordered list is being considered. This plot has been truncated at $x=0.4$ as, after this point, all curves vary minimally about 0, with two figures shown for clarity when comparing between curves. It can be seen that, when using the wavelet features, improvements in recall of up to 12 percentage points can be seen for both anomaly detection algorithms, whilst the \textit{feets} features improve only moderately. When combining \textit{feets} and wavelet features, there is a 20 percentage point increase when using the LOF, whilst the IF improvement is similar to that of each feature set alone.}
    \label{fig:residuals}
\end{figure*}

To quantify how much active learning changes our results, we subtract each updated recall curve from those in section \ref{sec:anomaly-results}. The residual curves for all feature-model pairs can be seen in Figure \ref{fig:residuals}. We can immediately see that the wavelet feature instances improve  increasing recall by as much as 10 per cent within the most anomalous region of feature space. Active learning applied to the \textit{feets} features improves anomaly recall for the most anomalous sources, but once $\sim$15 per cent of the data have been analysed, minimal improvement is seen. This is reflected in the respective $R_{10}$ and $R_{30}$ values seen in Table \ref{tab:recalls} which remain almost unchanged to two decimal places. The combined feature set shows a dramatic increase in performance with the LOF algorithm, by up to 20 percentage points at times. This corresponds to recovering approximately half of the transients in 8 per cent of the data compared to only around one third beforehand. By contrast, the combined feature set as applied to the IF shows approximately the same qualitative improvements as when applying the feature sets alone. As seen in the previous section, $(\eta, V)$ show no improvement and in some cases perform worse than before. This is in part because our active learning procedure involves labelling the sources at the top of our anomaly score-ordered list. Given that the anomaly detection performance using the standard $(\eta, V)$ feature pair essentially matches random chance and that transients make up $\sim$2 per cent of our sample, labelling only 2 per cent of the data will not provide the regressor with many positive anomalies from which to learn.

The precision values of Table \ref{tab:recalls} show minimal improvement following active learning. Only the $P_{10}$ values improve by, at most, 1 per cent, whilst all $P_{30}$ values remain the same to two decimal places. This is in part due to the asymptotic behaviour of these precision calculations as mentioned earlier, namely that as $n$ reaches our total sample size, the precision must approach the overall rate of transients in our data of $\sim$ 2 percent. As a result, it is hard to improve our $P_{30}$ values much when there are so few anomalies to find in our dataset.

In terms of AUC values, we see minimal change compared to the default anomaly detection procedure, with all \textit{feets} and $(\eta, V)$ values varying at the sub-per cent level and wavelet instances improving by only 0.01. This matches what we see in Figure \ref{fig:residuals} where after x=0.3 all curves vary only minimally around no change (y=0). As before we stress that this metric does not match the user-driven experience of finding anomalies preferentially in as small a data volume as possible. As before, the feets-LOF iteration provide the most transients in 10 per cent of the data, though the wavelet instances perform almost as well following the active learning. 

\section{Discussion}
\label{sec:disc}

\subsection{Feature performance and comparison}
In this work we have demonstrated the first application of anomaly detection to the task of finding transients with radio interferometers. In general we see that appropriate features, models and domain knowledge via active learning allow us to recover over half of our volunteer-verified transients (anomalies) in one tenth of the total data volume. Similarly, we have shown that the number of sources a human must inspect in order to find transients can as reduced by a factor of 5 compared to baseline performance. The choice of model used makes a small but noticeable difference, with the LOF algorithm generally performing better. The choice of feature or representation is a much stronger determinant of which and how many anomalies can be recovered. For example, when dealing with the large and heterogeneous sample of data used in this work, it is clear that a simple 2-parameter setup fails to represent the data in a way that allows the subsequent anomaly detection models to succeed. This is not a surprise, given that $\eta$ and $V$ are intended for intra-field comparisons i.e. between light curves in a single set of observations. Indeed these are still useful statistics that have aided in the discovery in numerous transients from the ThunderKAT observations (see section \ref{sec:intro}).

On the other hand, the \textit{feets} features show good promise at recovering transients and can recall over half of our anomalies in less than 10 per cent of the data. Our wavelet features, after active learning, perform almost as well as the \textit{feets} features with respect to $R_{10}$ and are by a significant margin the best feature set when comparing $R_{30}$ values. The large improvement seen in the wavelet features may be related to the clear clustering of anomalies seen in Figure \ref{fig:UMAP-WV}, where the majority of our transients are located in a relatively small part of this representation. Therefore we might conclude that if active learning adequately highlights sources from this region of feature space, performance will increase.

The combined feature set performs similar to both individual sets. 
The recall plot of the combined features is effectively an average of the \textit{feets} and wavelet feature sets. This implies that while both feature sets contribute information, they are not entirely orthogonal. The additional features will make some sources appear more anomalous while others will now have more in common with the larger population. The base anomaly detection algorithms are not able to make effective use of the additional features. However, applying active learning with the LOF dramatically improves performance of these features, improving recall by up to 20 percentage points. By using this training data, active learning can far more effectively take advantage of the additional information provided, leveraging either the \textit{feets} features or the wavelets depending on the region of feature space in which the labelled source is found. To avoid these differences caused by the initial anomaly detection algorithm, future active learning procedures may skip this step entirely \citep{Lochner2025}.

\begin{figure}
	\includegraphics[width=\columnwidth]{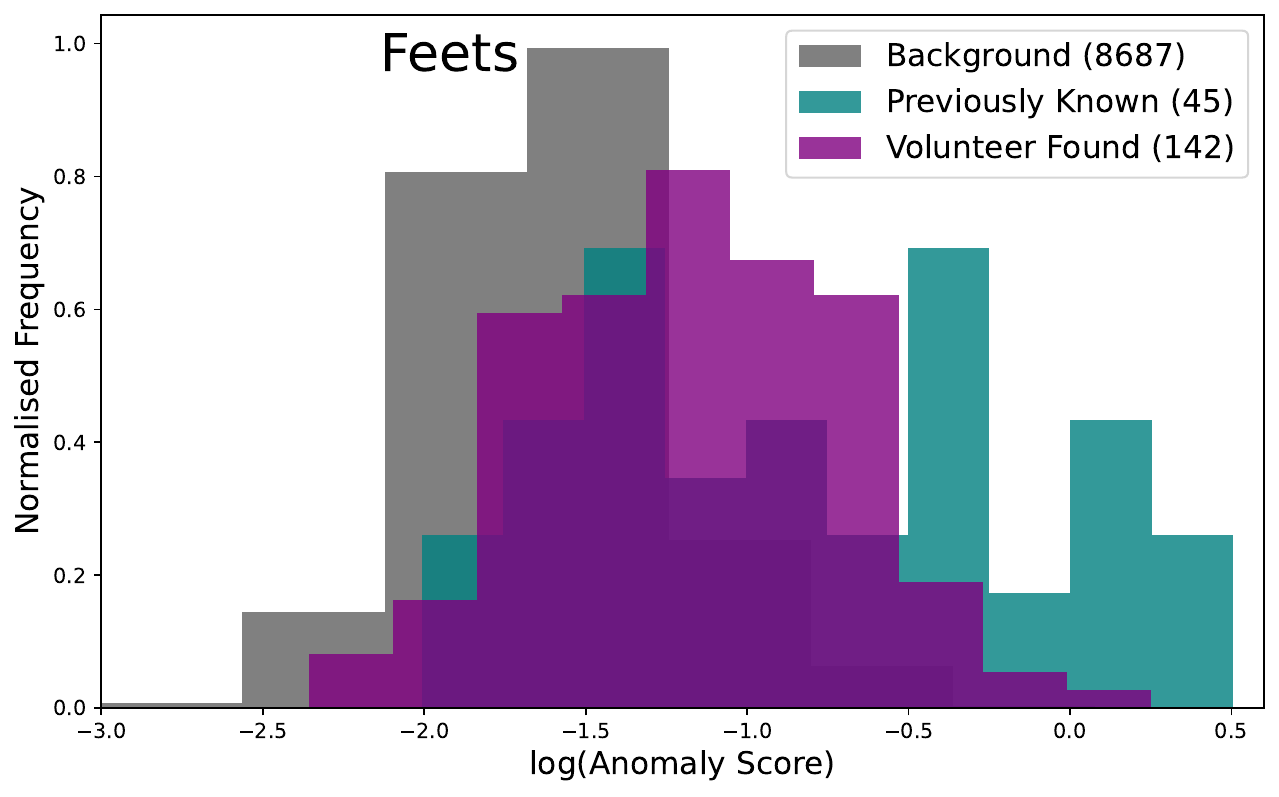}
 \includegraphics[width=\columnwidth]{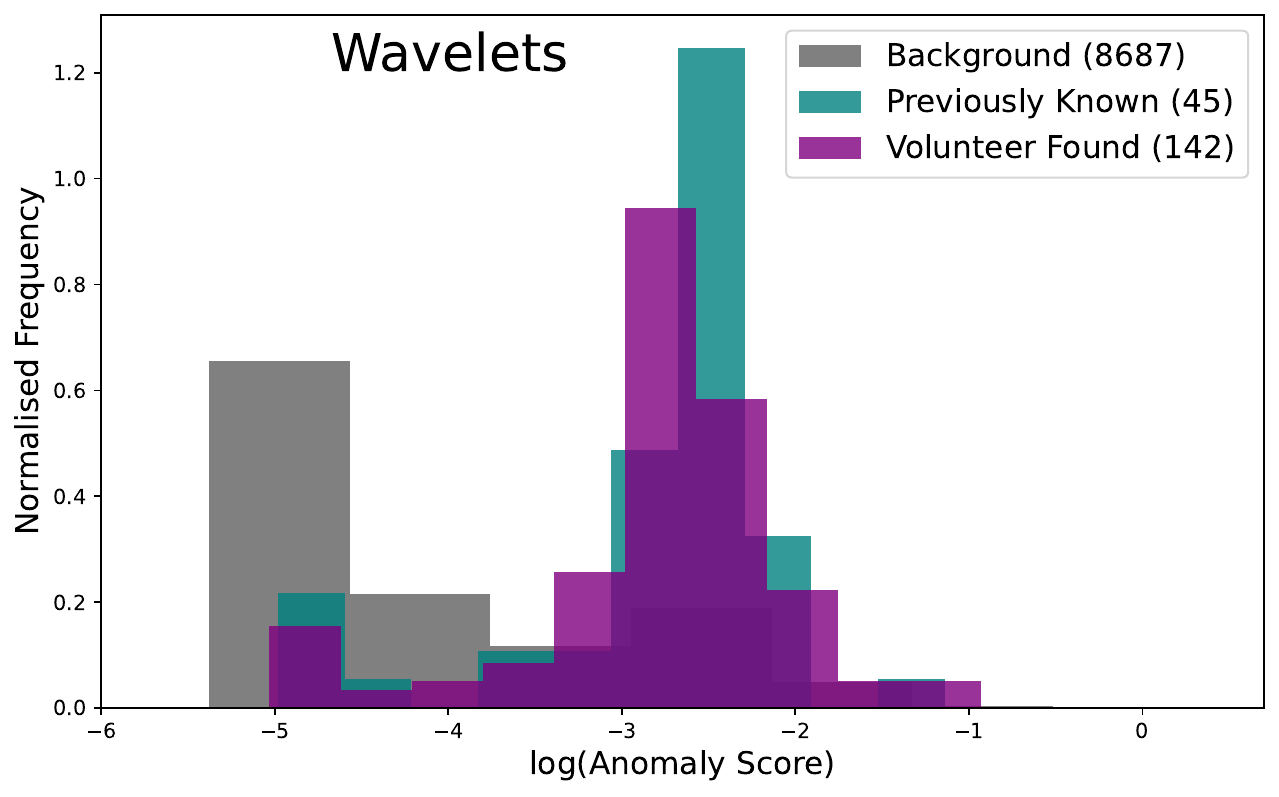}
 \caption{The distributions of the output scores of LOF with active learning, as applied to the \textit{feets} features (upper) and wavelet features (lower), with histograms normalised by area. Volunteer found sources are those uniquely discovered by citizen scientists in \citetalias{Andersson2023}, whilst those labelled as previously known are those that had been already identified by other works. All 8687 remaining sources are labelled as background.}
    \label{fig:distributions}
\end{figure}

A comparison between the \textit{feets} and wavelet features can be seen in Figure \ref{fig:distributions}, where the distributions of actively learnt scores between different classes are shown. In both instances, background (i.e. non-anomalous) sources receive lower anomaly scoures as expected. However, when comparing between previously known transients (45 sources, see \citetalias{Andersson2023} for the listed XRBs and other commensal discoveries) and those found by volunteers (142 sources), it can be seen that when using the wavelet features, these two classes produce similar, tightly peaked distributions in score i.e. they are approximately equally anomalous. However, when using the \textit{feets} features, the distribution of both desired classes are broader, with volunteer-found sources overlapping slightly more with background objects. This implies that, if searching only for sources such as the previously known XRBs, that the \textit{feets} features may be the feature set to use, but that the wavelet features can find both classes of transient with equal aptitude.  

We note that one could spend a great deal longer curating the perfect feature set for a given task and there is no guarantee that the features tested herein form the optimal choices. For example, we know that the heterogeneity of our time series and their sampling, along with the imputation of shorter light curves, likely induces biases in the wavelet calculations. However, without labels it would not be possible to tune hyperparameters in any stage of the process.
Nevertheless, that pre-existing methods work well in this regime is a significant step towards finding radio transients in real-time surveys. Future work may investigate whether, for example, autoencoder architectures (e.g. \citealp{Villar2021}) provide a way to extract meaningful features directly from the radio time series.

As a final point on feature selection, we have used Gaussian Processes only for interpolating light curves, though they serve as powerful tools in their own right. When fitting a GP, the covariance kernels used have hyperparameters that can provide alternative metrics for detecting variables and transients. This idea is explored further in Fu et al. (\textit{in prep.}), who use a multi-term covariance kernel and its hyperparameters to find candidate transients in the same data used here. This study finds that the joint distribution of hyperparameters is a good discriminant for separating variable and transient candidates from stable ones and goes on to detect transients missed by traditional metrics and by \citetalias{Andersson2023}.

\subsection{Computational Scalability}

With regards to scalability, all processing has been performed on a standard CPU  on a desktop with no attempt at optimisation. Despite this, all feature extraction and anomaly detection took less than 10 minutes, with the Gaussian process regression being the rate-limiting step. Even when considering the additional step of producing light curves from images in the first place, the overall run time is low enough to allow for a low-latency online transient detection system. Commensal real-time searches already exist on some interferometers, such as the \textit{realfast} system on the Karl G Jansky's Very Large Array \citep{Law2018} and on the Murchison Widefield Array \citep[see Methods in][]{Hurley-Walker2023}. This work shows that the incorporation of anomaly detection techniques would be a flexible addition to such a backend system, whereby the anomaly ranking and user-defined preference would allow for rapid discovery without ballooning data rates.

However, additional steps will be required in order to maximise potential discoveries from current and upcoming data streams. For example, several recent discoveries have been made using fast images (that is, sub-integrations of a single long observation) that warrant further investigation \citep{Hurley-Walker2022, Caleb2022, YWang2023}, while in this work the images are typically separated by a week. We plan to apply these anomaly detection methods to sets of 8s-images taken by MeerKAT in the near future. Similarly, we have not made use of any spectral or polarimetric information inherent in the data, both of which can provide crucial additional information and reduce contaminants \citep{Heywood2023}. However, it is impossible to know on which timescales and at which frequencies to search in a manner that is completely agnostic to all the possible underlying astrophysics. Therefore it is worth thinking about the specific science case in mind and how to best leverage these data products - for example, if one is searching for radio flaring stars, the use of Stokes V imaging is a powerful tool that eliminates many contaminants from images (e.g. \citealp{Pritchard2021, Pritchard2023}). These additional pipeline steps would, however, add to the computational time of finding these anomalies. It is therefore down to the given science case whether data rates are such that only the bare minimum information can be extracted, or whether the astrophysical phenomena necessitate additional processing.

\subsection{Citizen Science}

\begin{figure}
	\includegraphics[width=\columnwidth]{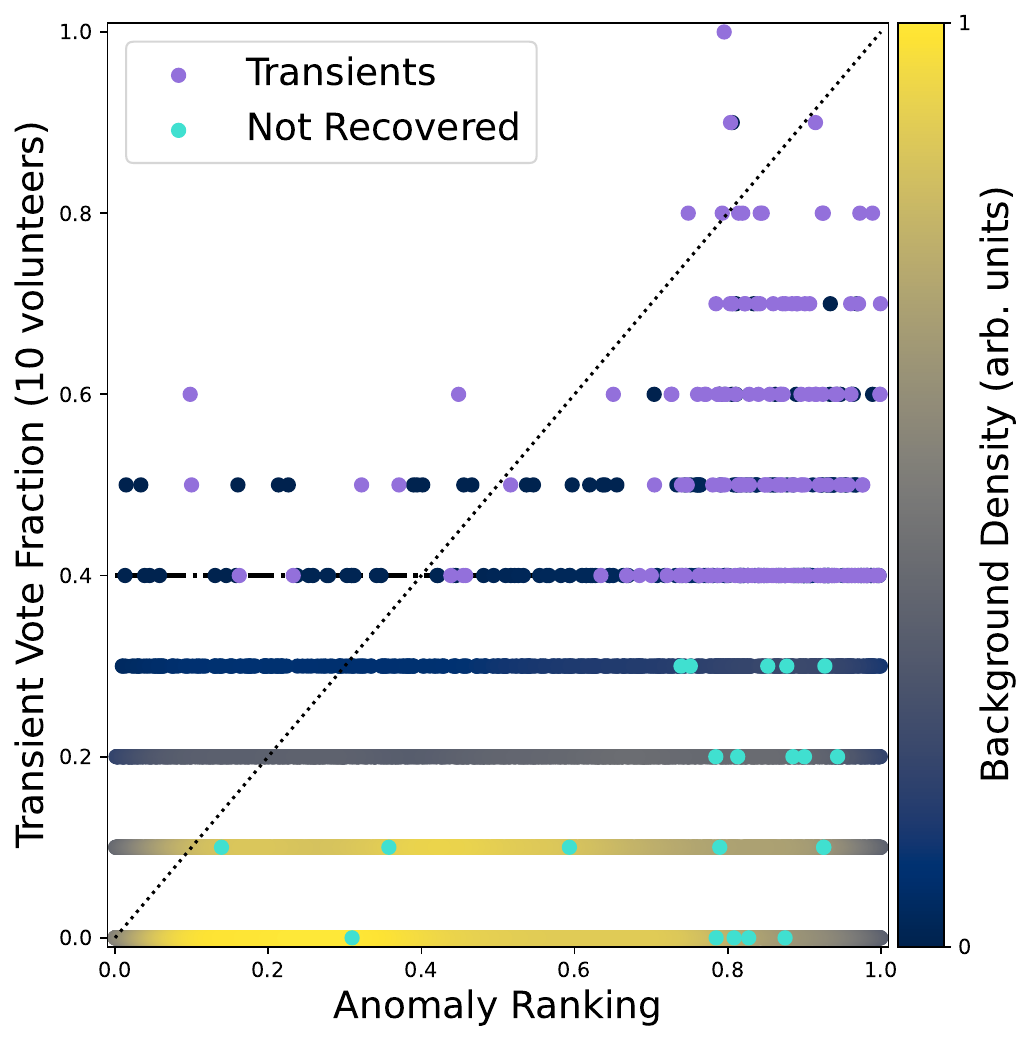}
    \caption{A comparison between how many citizen scientists voted for a source as a transient and its ranking in the anomaly detection work, where a rank of 1 indicates the most anomalous source, whilst the least anomalous source has a ranking of 0. Sources ‘not recovered' refers to known variables/transients in the light curves that fall below the 4/10 threshold of \citetalias{Andersson2023} and were therefore regarded as missed by volunteers. The background points are all remaining sources, coloured by the density of points. The anomaly list used here corresponds to the wavelet features applied to the LOF algorithm, but results are consistent for all feature-model pairs.}
    \label{fig:volunteers}
\end{figure}

All of the discoveries used as our ground truth were found by volunteers. With upcoming (big) data projects such as the SKA and LSST, it is worth discussing the role of individuals and groups with regards to discovery science. A crude but immediate comparison between human discoveries and automated anomaly detection algorithms can be seen in Figure \ref{fig:volunteers}, where the transient vote fraction is how many out of 10 volunteers that saw each source voted for it as a transient/variable. The anomaly ranking is a source's position in the anomaly score-ordered list for the wavelet-LOF instance, though other feature-model pairs were tested with qualitatively similar results. We can see that the majority of volunteer transients are ranked highly by the anomaly detection model, matching the recall curves seen earlier. However, we can also see that there are known transients and variables missed by volunteers (i.e below \citetalias{Andersson2023}'s threshold of 0.4) - these are typically low significance sources and/or those that show variability that was difficult for volunteers to identify but were found in previous studies i.e. \citet{Driessen2020, Rowlinson2022, Driessen2022}. It is encouraging then that the anomaly detection methods rank these anomalies as highly as many of the new transients/variables confirmed by volunteers, indicating that we can avoid a potential source of bias in our searches. Similarly, there are sources given low scores by the anomaly detection algorithms (even with active learning) that were clearly identified by a majority of volunteers - these are the data points about the $y=x$ line in Figure \ref{fig:volunteers}. So these are anomalies that volunteers find "more easily" than our algorithms. It is therefore pertinent to leverage each search method in a complementary manner \citep[see also][]{Wright2017}. For example, one could perform a first pass with anomaly detection procedures, which would find the highly anomalous sources easily and then pass the sources that are not ranked highly (e.g. beyond the ‘knee' of a given recall curve) to citizen scientists. However, it is also important to value the time taken by individual volunteers and consider whether asking them to find relatively few sources in large data volumes is an optimal or even fair division of labour. We caveat that this brief comparison with volunteer performance may not necessarily generalise to other projects, future data releases or other volunteers and that project builders could in theory give different training and instructions, producing different results. Nevertheless, the combination of rapid computation and the advantages of volunteer-led searches in concert with expert analysis and, as demonstrated in this work, anomaly detection techniques, hold great promise for future transient hunting efforts.

Involving citizen scientists at the active learning stage is one way to leverage this combination of algorithmic efficiency and volunteer diligence. As simulated and then demonstrated by the \textit{Galaxy Zoo} team \citep{Walmsley2020MNRAS.491.1554W, Walmsley2022a}, volunteers can be preferentially shown sources which are most informative for a model to train on, dramatically reducing the number of votes required overall. In an unsupervised setting this human-in-the-loop cycle might constitute showing volunteers the most anomalous sources after default anomaly detection and then updating an acquisition function based on the citizen scientists' classifications. Volunteers could then be shown sources according to the new anomaly score-ordered list and the process repeats. Due to the subjective nature of what a given (citizen) scientist might find interesting, the resulting acquisition function would be some average of many different opinions, weighted by the number of sources classified per volunteer. Project scientists would then have to decide whether or not to intentionally attempt to influence behaviour by, for example, displaying 'wanted' source classes in volunteer training, or whether a diversity in voting patterns can be leveraged for downstream tasks. In any case, optimising what sources are shown to volunteers will be of paramount importance when dealing with future surveys such as from the SKA or Vera C. Rubin Observatory.

\section{Conclusions}
\label{sec:conclude}

In this paper we have presented the first analysis of anomaly detection models and active learning for finding radio transients via their light curves. We are able to make use of volunteer contributions on the \textit{Bursts from Space: MeerKAT} citizen science project as a ground truth sample against which to test these unsupervised machine learning techniques incorporated into the \textsc{Astronomaly} package. We have shown how choices in feature sets result in different clustering of sources and how the appropriate representation of data has a strong impact on our ability to recover anomalies. We have also shown that choice of anomaly detection model is a less significant determinant in performance. For any combination of feature set and model, that the hit-rate of transients is always at least a factor of two greater than previous methods, for highly anomalous sources. Active learning on a small subset of the data produces  better recall of transients, particularly for those ranked as highly anomalous. This enables users to focus on anomalies that are interesting for their given science case. These promising methods, in combination with the volunteer-led efforts and commensal search engines being built into telescope backends will allow for unprecedented discovery of rare and novel transients that will drive science cases for upcoming facilities such as the SKA.

\section*{Acknowledgements}

AA acknowledges the support given by the Science and Technology Facilities Council through a STFC studentship and the support of the Breakthrough Listen project. Breakthrough Listen is managed by the Breakthrough Initiatives, sponsored by the Breakthrough Prize Foundation. 
CJL acknowledges support from the Alfred P. Sloan foundation.
ML acknowledges support from the South African Radio Astronomy Observatory and the National Research Foundation (NRF) towards this research. Opinions expressed and conclusions arrived at, are those of the authors and are not necessarily to be attributed to the NRF.
JvdE acknowledges a Warwick Astrophysics prize post-doctoral fellowship made possible thanks to a generous philanthropic donation, and was supported by funding from the European Union's Horizon Europe research and innovation programme under the Marie Skłodowska-Curie grant agreement No. 101148693 (MeerSHOCKS) for part of this work.
MV acknowledges financial support from the Inter-University Institute for Data Intensive Astronomy (IDIA), a partnership of the University of Cape Town, the University of Pretoria and the University of the Western Cape, and from the South African Department of Science and Innovation's National Research Foundation under the ISARP RADIOMAP Joint Research Scheme (DSI-NRF Grant Number 150551) and the CPRR HIPPO Project (DSI-NRF Grant Number SRUG22031677).

This publication uses data generated via the \url{Zooniverse.org} platform, development of which is funded by generous support, including a Global Impact Award from Google, and by a grant from the Alfred P. Sloan Foundation.

MeerKAT is operated by the South African Radio Astronomy Observatory (SARAO), which is a facility of the National Research Foundation, an agency of the Department of Science and Innovation.  We thank the SARAO staff involved in obtaining the MeerKAT observations.

We acknowledge the use of the ilifu cloud computing facility – \url{www.ilifu.ac.za}, a partnership between the University of Cape Town, the University of the Western Cape, Stellenbosch University, Sol Plaatje University and the Cape Peninsula University of Technology. The ilifu facility is supported by contributions from the Inter-University Institute for Data Intensive Astronomy (IDIA – a partnership between the University of Cape Town, the University of Pretoria and the University of the Western Cape), the Computational Biology division at UCT and the Data Intensive Research Initiative of South Africa (DIRISA).
This work made use of the CARTA (Cube Analysis and Rendering Tool for Astronomy) software (DOI \href{https://zenodo.org/records/4905459}{\texttt{10.5281/zenodo.3377984}} –  \url{https://cartavis.github.io}).

This research has made use of the Python programming language and the following open source packages: Astropy \citep{Astropy2013, Astropy2018, Astropy2022}, Matplotlib \citep{matplotlib2007}, NumPy \citep{Harris2020Numpy}, Pandas \citep{mckinney-pandas-2010}, PyWavelets \citep{Lee2019Wavelets}, scikit-learn \citep{Pedregosa2011}, SciPy \citep{2020SciPy-NMeth} and umap-learn \citep{McInnes2018,sainburg202UMAP}.

\section*{Data Availability}

ThunderKAT raw data are available on the SARAO archive (\url{https://archive.sarao.ac.za/}). Light curves of all data used herein can be found at \url{https://github.com/AnderssonAstro/BfS-MKT-Analysis} with digital object identifier \href{https://zenodo.org/records/10462688}{\texttt{10.5281/zenodo.10462687}}. 

All codebases used for feature extraction, anomaly detection and active learning are open source and can be found from \citet{Cabral2018ascl, Cabral2018}, \citet{Alves2022}, \citet{Lochner2021} and \citet{Pedregosa2011} as appropriate.



\bibliographystyle{mnras}
\bibliography{PaperDraft1.bib} 

\begin{thebibliography}{}
\makeatletter
\relax
\def\mn@urlcharsother{\let\do\@makeother \do\$\do\&\do\#\do\^\do\_\do\%\do\~}
\def\mn@doi{\begingroup\mn@urlcharsother \@ifnextchar [ {\mn@doi@} {\mn@doi@[]}}
\def\mn@doi@[#1]#2{\def\@tempa{#1}\ifx\@tempa\@empty \href {http://dx.doi.org/#2} {doi:#2}\else \href {http://dx.doi.org/#2} {#1}\fi \endgroup}
\def\mn@eprint#1#2{\mn@eprint@#1:#2::\@nil}
\def\mn@eprint@arXiv#1{\href {http://arxiv.org/abs/#1} {{\tt arXiv:#1}}}
\def\mn@eprint@dblp#1{\href {http://dblp.uni-trier.de/rec/bibtex/#1.xml} {dblp:#1}}
\def\mn@eprint@#1:#2:#3:#4\@nil{\def\@tempa {#1}\def\@tempb {#2}\def\@tempc {#3}\ifx \@tempc \@empty \let \@tempc \@tempb \let \@tempb \@tempa \fi \ifx \@tempb \@empty \def\@tempb {arXiv}\fi \@ifundefined {mn@eprint@\@tempb}{\@tempb:\@tempc}{\expandafter \expandafter \csname mn@eprint@\@tempb\endcsname \expandafter{\@tempc}}}

\bibitem[\protect\citeauthoryear{Alves, Peiris, Lochner, McEwen, Allam  \& Biswas}{Alves et~al.}{2022}]{Alves2022}
Alves C.~S.,  Peiris H.~V.,  Lochner M.,  McEwen J.~D.,  Allam T.,   Biswas R.,  2022, \mn@doi [\apjs] {10.3847/1538-4365/ac3479}, 258, 23

\bibitem[\protect\citeauthoryear{Andersson et~al.,}{Andersson et~al.}{2022}]{Andersson2022}
Andersson A.,  et~al., 2022, \mn@doi [\mnras] {10.1093/mnras/stac1002}, 513, 3482

\bibitem[\protect\citeauthoryear{Andersson et~al.,}{Andersson et~al.}{2023}]{Andersson2023}
Andersson A.,  et~al., 2023, \mn@doi [\mnras] {10.1093/mnras/stad1298}, 523, 2219

\bibitem[\protect\citeauthoryear{Arfaoui, Mabrouk  \& Cattani}{Arfaoui et~al.}{2021}]{arfaoui2021wavelet}
Arfaoui S.,  Mabrouk A.,   Cattani C.,  2021, Wavelet Analysis: Basic Concepts and Applications.
CRC Press, \url {https://books.google.co.uk/books?id=ZT4oEAAAQBAJ}

\bibitem[\protect\citeauthoryear{{Braun}, {Bonaldi}, {Bourke}, {Keane}  \& {Wagg}}{{Braun} et~al.}{2019}]{Braun2019arXiv191212699B}
{Braun} R.,  {Bonaldi} A.,  {Bourke} T.,  {Keane} E.,   {Wagg} J.,  2019, \mn@doi [arXiv e-prints] {10.48550/arXiv.1912.12699}, \href {https://ui.adsabs.harvard.edu/abs/2019arXiv191212699B} {p. arXiv:1912.12699}

\bibitem[\protect\citeauthoryear{Breiman}{Breiman}{2001}]{Breiman2001}
Breiman L.,  2001, \mn@doi [Machine Learning] {10.1023/A:1010933404324}, 45, 5

\bibitem[\protect\citeauthoryear{Breuniq, Kriegel, Ng  \& Sander}{Breuniq et~al.}{2000}]{Breuniq2000}
Breuniq M.~M.,  Kriegel H.~P.,  Ng R.~T.,   Sander J.,  2000, \mn@doi [ACM SIGMOD Record] {10.1145/335191.335388}, 29, 93

\bibitem[\protect\citeauthoryear{{Bright} et~al.,}{{Bright} et~al.}{2020}]{Bright2020NatAs...4..697B}
{Bright} J.~S.,  et~al., 2020, \mn@doi [Nature Astronomy] {10.1038/s41550-020-1023-5}, \href {https://ui.adsabs.harvard.edu/abs/2020NatAs...4..697B} {4, 697}

\bibitem[\protect\citeauthoryear{{Byrd}, {Lu}, {Nocedal}  \& {Zhu}}{{Byrd} et~al.}{1995}]{Byrd1995SJSC...16.1190B}
{Byrd} R.~H.,  {Lu} P.,  {Nocedal} J.,   {Zhu} C.,  1995, \mn@doi [SIAM Journal on Scientific Computing] {10.1137/0916069}, \href {https://ui.adsabs.harvard.edu/abs/1995SJSC...16.1190B} {16, 1190}

\bibitem[\protect\citeauthoryear{{Cabral}, {Sanchez}, {Ramos}, {Gurovich}, {Granitto}  \& {VanderPlas}}{{Cabral} et~al.}{2018a}]{Cabral2018ascl}
{Cabral} J.~B.,  {Sanchez} B.,  {Ramos} F.,  {Gurovich} S.,  {Granitto} P.,   {VanderPlas} J.,  2018a, Astrophysics Source Code Library, p. ascl:1806.001

\bibitem[\protect\citeauthoryear{Cabral, S{\'{a}}nchez, Ramos, Gurovich, Granitto  \& Vanderplas}{Cabral et~al.}{2018b}]{Cabral2018}
Cabral J.~B.,  S{\'{a}}nchez B.,  Ramos F.,  Gurovich S.,  Granitto P.~M.,   Vanderplas J.,  2018b, \mn@doi [Astronomy and Computing] {10.1016/j.ascom.2018.09.005}, 25, 213

\bibitem[\protect\citeauthoryear{Caleb et~al.,}{Caleb et~al.}{2022}]{Caleb2022}
Caleb M.,  et~al., 2022, \mn@doi [Nature Astronomy] {10.1038/s41550-022-01688-x}, 6, 828

\bibitem[\protect\citeauthoryear{Caleb et~al.,}{Caleb et~al.}{2024}]{Caleb2024NatAs...8.1159C}
Caleb M.,  et~al., 2024, \mn@doi [Nature Astronomy] {10.1038/s41550-024-02277-w}, \href {https://ui.adsabs.harvard.edu/abs/2024NatAs...8.1159C} {8, 1159}

\bibitem[\protect\citeauthoryear{{Chastain}, {van der Horst}, {Rowlinson}, {Rhodes}, {Andersson}, {Diretse}, {Fender}  \& {Woudt}}{{Chastain} et~al.}{2023}]{Chastain2023}
{Chastain} S.~I.,  {van der Horst} A.~J.,  {Rowlinson} A.,  {Rhodes} L.,  {Andersson} A.,  {Diretse} R.,  {Fender} R.~P.,   {Woudt} P.~A.,  2023, \mn@doi [\mnras] {10.1093/mnras/stad2714}, \href {https://ui.adsabs.harvard.edu/abs/2023MNRAS.526.1888C} {526, 1888}

\bibitem[\protect\citeauthoryear{{Chastain} et~al.,}{{Chastain} et~al.}{2024}]{Chastiain2025}
{Chastain} S.~I.,  et~al., 2024, \mn@doi [arXiv e-prints] {10.48550/arXiv.2412.02832}, \href {https://ui.adsabs.harvard.edu/abs/2024arXiv241202832C} {p. arXiv:2412.02832}

\bibitem[\protect\citeauthoryear{Czech, Mishra  \& Inggs}{Czech et~al.}{2018}]{Czech2018}
Czech D.,  Mishra A.,   Inggs M.,  2018, Astronomy and Computing, 25, 52

\bibitem[\protect\citeauthoryear{{Debosscher}, {Sarro}, {Aerts}, {Cuypers}, {Vandenbussche}, {Garrido}  \& {Solano}}{{Debosscher} et~al.}{2007}]{Debosscher2007}
{Debosscher} J.,  {Sarro} L.~M.,  {Aerts} C.,  {Cuypers} J.,  {Vandenbussche} B.,  {Garrido} R.,   {Solano} E.,  2007, \mn@doi [Astronomy and Astrophysics] {10.1051/0004-6361:20077638}, 475, 1159

\bibitem[\protect\citeauthoryear{Driessen et~al.,}{Driessen et~al.}{2020}]{Driessen2020}
Driessen L.~N.,  et~al., 2020, \mn@doi [\mnras] {10.1093/mnras/stz3027}, 491, 560

\bibitem[\protect\citeauthoryear{Driessen, Williams, McDonald, Stappers, Buckley, Fender  \& Woudt}{Driessen et~al.}{2022a}]{Driessen2021}
Driessen L.~N.,  Williams D.~R.,  McDonald I.,  Stappers B.~W.,  Buckley D.~A.,  Fender R.~P.,   Woudt P.~A.,  2022a, \mn@doi [\mnras] {10.1093/mnras/stab3461}, 510, 1083

\bibitem[\protect\citeauthoryear{Driessen et~al.,}{Driessen et~al.}{2022b}]{Driessen2022}
Driessen L.~N.,  et~al., 2022b, \mn@doi [\mnras] {10.1093/mnras/stac756}, 512, 5037

\bibitem[\protect\citeauthoryear{{Driessen} et~al.,}{{Driessen} et~al.}{2024}]{Driessen2024}
{Driessen} L.~N.,  et~al., 2024, \mn@doi [\mnras] {10.1093/mnras/stad3329}, \href {https://ui.adsabs.harvard.edu/abs/2024MNRAS.527.3659D} {527, 3659}

\bibitem[\protect\citeauthoryear{{Etsebeth}, {Lochner}, {Walmsley}  \& {Grespan}}{{Etsebeth} et~al.}{2024}]{Etsebeth2023}
{Etsebeth} V.,  {Lochner} M.,  {Walmsley} M.,   {Grespan} M.,  2024, \mn@doi [\mnras] {10.1093/mnras/stae496}, \href {https://ui.adsabs.harvard.edu/abs/2024MNRAS.529..732E} {529, 732}

\bibitem[\protect\citeauthoryear{Fender et~al.,}{Fender et~al.}{2016}]{Fender2016}
Fender R.,  et~al., 2016, in Proceedings of Science.  (\mn@eprint {arXiv} {1711.04132}), \mn@doi{10.22323/1.277.0013}, \url {http://pos.sissa.it/}

\bibitem[\protect\citeauthoryear{{Fijma} et~al.,}{{Fijma} et~al.}{2024}]{Fijma2024}
{Fijma} S.,  et~al., 2024, \mn@doi [\mnras] {10.1093/mnras/stae382}, \href {https://ui.adsabs.harvard.edu/abs/2024MNRAS.528.6985F} {528, 6985}

\bibitem[\protect\citeauthoryear{Giles \& Walkowicz}{Giles \& Walkowicz}{2019}]{Giles2019}
Giles D.,  Walkowicz L.,  2019, \mn@doi [\mnras] {10.1093/mnras/sty3461}, 484, 834

\bibitem[\protect\citeauthoryear{Harris et~al.,}{Harris et~al.}{2020}]{Harris2020Numpy}
Harris C.~R.,  et~al., 2020, \mn@doi [\nat] {10.1038/s41586-020-2649-2}, 585, 357

\bibitem[\protect\citeauthoryear{Heywood}{Heywood}{2020}]{Heywood2020}
Heywood I.,  2020, ASCL, p. ascl:2009.003

\bibitem[\protect\citeauthoryear{Heywood}{Heywood}{2023}]{Heywood2023}
Heywood I.,  2023, \mn@doi [MNRAS] {10.1093/mnrasl/slad094}, 525, L76

\bibitem[\protect\citeauthoryear{{Heywood} et~al.,}{{Heywood} et~al.}{2022}]{Heywood2022}
{Heywood} I.,  et~al., 2022, \mn@doi [\mnras] {10.1093/mnras/stab3021}, \href {https://ui.adsabs.harvard.edu/abs/2022MNRAS.509.2150H} {509, 2150}

\bibitem[\protect\citeauthoryear{Hotan et~al.,}{Hotan et~al.}{2021}]{Hotan2021}
Hotan A.~W.,  et~al., 2021, \mn@doi [\pasa] {10.1017/pasa.2021.1}, 38

\bibitem[\protect\citeauthoryear{{Hughes}, {Aller}, {Aller}  \& {}}{{Hughes} et~al.}{1992}]{Hughes1992}
{Hughes} P.~A.,  {Aller} H.~D.,  {Aller} M.~F.,   {} 1992, \mn@doi [The Astrophysical Journal] {10.1086/171734}, 396, 469

\bibitem[\protect\citeauthoryear{{Hunter}}{{Hunter}}{2007}]{matplotlib2007}
{Hunter} J.~D.,  2007, \mn@doi [Computing in Science and Engineering] {10.1109/MCSE.2007.55}, \href {https://ui.adsabs.harvard.edu/abs/2007CSE.....9...90H} {9, 90}

\bibitem[\protect\citeauthoryear{Hurley-Walker et~al.,}{Hurley-Walker et~al.}{2022}]{Hurley-Walker2022}
Hurley-Walker N.,  et~al., 2022, \mn@doi [\nat] {10.1038/s41586-021-04272-x}, 601, 526

\bibitem[\protect\citeauthoryear{Hurley-Walker et~al.,}{Hurley-Walker et~al.}{2023}]{Hurley-Walker2023}
Hurley-Walker N.,  et~al., 2023, \mn@doi [\nat] {10.1038/s41586-023-06202-5}, 619, 487

\bibitem[\protect\citeauthoryear{{Hurley-Walker} et~al.,}{{Hurley-Walker} et~al.}{2024}]{HurleyWalker2024}
{Hurley-Walker} N.,  et~al., 2024, \mn@doi [\apjl] {10.3847/2041-8213/ad890e}, \href {https://ui.adsabs.harvard.edu/abs/2024ApJ...976L..21H} {976, L21}

\bibitem[\protect\citeauthoryear{Ishida et~al.,}{Ishida et~al.}{2021}]{Ishida2021}
Ishida E.~E.,  et~al., 2021, \mn@doi [\aap] {10.1051/0004-6361/202037709}, 650, A195

\bibitem[\protect\citeauthoryear{Ivezi{\'{c}} et~al.,}{Ivezi{\'{c}} et~al.}{2019}]{Ivezic2019}
Ivezi{\'{c}} {\v{Z}}.,  et~al., 2019, \mn@doi [\apj] {10.3847/1538-4357/ab042c}, 873, 111

\bibitem[\protect\citeauthoryear{{Jankowski} et~al.,}{{Jankowski} et~al.}{2022}]{Jankowski2022}
{Jankowski} F.,  et~al., 2022, in {Ruiz} J.~E.,  {Pierfedereci} F.,   {Teuben} P.,  eds,  Astronomical Society of the Pacific Conference Series Vol. 532, Astronomical Society of the Pacific Conference Series. p.~273 (\mn@eprint {arXiv} {2012.05173}), \mn@doi{10.48550/arXiv.2012.05173}

\bibitem[\protect\citeauthoryear{Jonas \& {MeerKAT Team}}{Jonas \& {MeerKAT Team}}{2016}]{Jonas2016}
Jonas J.~L.,  {MeerKAT Team} T.,  2016, in Proceedings of MeerKAT Science: on the Pathway to the SKA. Sissa Medialab Srl, Stellenbosch, p.~1, \mn@doi{10.22323/1.277.0001}, \url {https://ui.adsabs.harvard.edu/abs/2016mks..confE...1J/abstract}

\bibitem[\protect\citeauthoryear{{Kim}, {Protopapas}, {Alcock}, {Byun}  \& {Bianco}}{{Kim} et~al.}{2009}]{Kim2009}
{Kim} D.-W.,  {Protopapas} P.,  {Alcock} C.,  {Byun} Y.-I.,   {Bianco} F.~B.,  2009, \mn@doi [Monthly Notices of the Royal Astronomical Society] {10.1111/j.1365-2966.2009.14967.x}, 397, 558

\bibitem[\protect\citeauthoryear{Kim et~al.,}{Kim et~al.}{2011}]{Kim2011}
Kim D.-W.,  et~al., 2011, \mn@doi [ApJ] {10.1088/0004-637X/735/2/68}, 735, 68

\bibitem[\protect\citeauthoryear{{Kim}, {Protopapas}, {Bailer-Jones}, {Byun}, {Chang}, {Marquette}  \& {Shin}}{{Kim} et~al.}{2014}]{Kim2014}
{Kim} D.-W.,  {Protopapas} P.,  {Bailer-Jones} C. A.~L.,  {Byun} Y.-I.,  {Chang} S.-W.,  {Marquette} J.-B.,   {Shin} M.-S.,  2014, \mn@doi [Astronomy and Astrophysics] {10.1051/0004-6361/201323252}, 566, A43

\bibitem[\protect\citeauthoryear{Law et~al.,}{Law et~al.}{2018}]{Law2018}
Law C.~J.,  et~al., 2018, \mn@doi [\apjs] {10.3847/1538-4365/aab77b}, 236, 8

\bibitem[\protect\citeauthoryear{Lee, Gommers, Waselewski, Wohlfahrt  \& OLeary}{Lee et~al.}{2019}]{Lee2019Wavelets}
Lee G.~R.,  Gommers R.,  Waselewski F.,  Wohlfahrt K.,   OLeary A.,  2019, \mn@doi [Journal of Open Source Software] {10.21105/joss.01237}, 4, 1237

\bibitem[\protect\citeauthoryear{Liu, Ting  \& Zhou}{Liu et~al.}{2008}]{Liu2008}
Liu F.~T.,  Ting K.~M.,   Zhou Z.~H.,  2008, \mn@doi [Proceedings - IEEE International Conference on Data Mining, ICDM] {10.1109/ICDM.2008.17}, pp 413--422

\bibitem[\protect\citeauthoryear{Lochner \& Bassett}{Lochner \& Bassett}{2021}]{Lochner2021}
Lochner M.,  Bassett B.~A.,  2021, \mn@doi [Astronomy and Computing] {10.1016/j.ascom.2021.100481}, 36

\bibitem[\protect\citeauthoryear{{Lochner} \& {Rudnick}}{{Lochner} \& {Rudnick}}{2025}]{Lochner2025}
{Lochner} M.,  {Rudnick} L.,  2025, \mn@doi [\aj] {10.3847/1538-3881/ada14c}, \href {https://ui.adsabs.harvard.edu/abs/2025AJ....169..121L} {169, 121}

\bibitem[\protect\citeauthoryear{{Lochner}, McEwen, Peiris, Lahav  \& Winter}{{Lochner} et~al.}{2016}]{Lochner2016}
{Lochner} M.,  McEwen J.~D.,  Peiris H.~V.,  Lahav O.,   Winter M.~K.,  2016, \mn@doi [\apjs] {10.3847/0067-0049/225/2/31}, 225, 31

\bibitem[\protect\citeauthoryear{{Lochner}, Rudnick, Heywood, Knowles  \& Shabala}{{Lochner} et~al.}{2023}]{Lochner2023}
{Lochner} M.,  Rudnick L.,  Heywood I.,  Knowles K.,   Shabala S.~S.,  2023, \mn@doi [\mnras] {10.1093/MNRAS/STAD074}, 000, 1

\bibitem[\protect\citeauthoryear{Lomb}{Lomb}{1976}]{Lomb1976}
Lomb N.~R.,  1976, \mn@doi [\apss] {10.1007/BF00648343}, 39, 447

\bibitem[\protect\citeauthoryear{Malanchev et~al.,}{Malanchev et~al.}{2021}]{Malanchev2021}
Malanchev K.~L.,  et~al., 2021, \mn@doi [\mnras] {10.1093/mnras/stab316}, 502, 5147

\bibitem[\protect\citeauthoryear{{Malenta} et~al.,}{{Malenta} et~al.}{2020}]{Malenta2020}
{Malenta} M.,  et~al., 2020, in {Pizzo} R.,  {Deul} E.~R.,  {Mol} J.~D.,  {de Plaa} J.,   {Verkouter} H.,  eds,  Astronomical Society of the Pacific Conference Series Vol. 527, Astronomical Data Analysis Software and Systems XXIX. p.~457

\bibitem[\protect\citeauthoryear{McInnes, Healy  \& Melville}{McInnes et~al.}{2018}]{McInnes2018}
McInnes L.,  Healy J.,   Melville J.,  2018, UMAP: Uniform Manifold Approximation and Projection for Dimension Reduction (\mn@eprint {arXiv} {1802.03426}), \url {https://arxiv.org/abs/1802.03426}

\bibitem[\protect\citeauthoryear{{M}c{K}inney}{{M}c{K}inney}{2010}]{mckinney-pandas-2010}
{M}c{K}inney W.,  2010, in {S}t\'efan van~der {W}alt {J}arrod {M}illman eds, {P}roceedings of the 9th {P}ython in {S}cience {C}onference. pp 56 -- 61, \mn@doi{10.25080/Majora-92bf1922-00a}

\bibitem[\protect\citeauthoryear{{Mesarcik}, {Boonstra}, {Iacobelli}, {Ranguelova}, {de Laat}  \& {van Nieuwpoort}}{{Mesarcik} et~al.}{2023}]{Mesarcik2023A&A...680A..74M}
{Mesarcik} M.,  {Boonstra} A.~J.,  {Iacobelli} M.,  {Ranguelova} E.,  {de Laat} C.~T.~A.~M.,   {van Nieuwpoort} R.~V.,  2023, \mn@doi [\aap] {10.1051/0004-6361/202347182}, \href {https://ui.adsabs.harvard.edu/abs/2023A&A...680A..74M} {680, A74}

\bibitem[\protect\citeauthoryear{{Mohale} \& {Lochner}}{{Mohale} \& {Lochner}}{2024}]{Mohale2023}
{Mohale} K.,  {Lochner} M.,  2024, \mn@doi [\mnras] {10.1093/mnras/stae926}, \href {https://ui.adsabs.harvard.edu/abs/2024MNRAS.530.1274M} {530, 1274}

\bibitem[\protect\citeauthoryear{Mooley et~al.,}{Mooley et~al.}{2016}]{Mooley2016}
Mooley K.~P.,  et~al., 2016, \mn@doi [\apj] {10.3847/0004-637x/818/2/105}, 818, 105

\bibitem[\protect\citeauthoryear{Murphy et~al.,}{Murphy et~al.}{2013}]{Murphy2013}
Murphy T.,  et~al., 2013, \mn@doi [\pasa] {10.1017/pasa.2012.006}, 30, e006

\bibitem[\protect\citeauthoryear{Murphy et~al.,}{Murphy et~al.}{2017}]{Murphy2017}
Murphy T.,  et~al., 2017, \mn@doi [\mnras] {10.1093/mnras/stw3087}, 466, 1944

\bibitem[\protect\citeauthoryear{Muthukrishna, Mandel, Lochner, Webb  \& Narayan}{Muthukrishna et~al.}{2022}]{Muthukrishna2021}
Muthukrishna D.,  Mandel K.~S.,  Lochner M.,  Webb S.,   Narayan G.,  2022, \mn@doi [\mnras] {10.1093/mnras/stac2582}, 517, 393

\bibitem[\protect\citeauthoryear{Nun et~al.,}{Nun et~al.}{2015}]{Nun2015}
Nun I.,  et~al., 2015, \mn@doi [arXiv] {10.48550/ARXIV.1506.00010}, p. arXiv:1506.00010

\bibitem[\protect\citeauthoryear{Pearson}{Pearson}{1901}]{Pearson1901}
Pearson K.,  1901, \mn@doi [The London, Edinburgh, and Dublin Philosophical Magazine and Journal of Science] {10.1080/14786440109462720}, 2, 559

\bibitem[\protect\citeauthoryear{{Pedregosa} et~al.,}{{Pedregosa} et~al.}{2011}]{Pedregosa2011}
{Pedregosa} F.,  et~al., 2011, \mn@doi [Journal of Machine Learning Research] {10.48550/arXiv.1201.0490}, \href {https://ui.adsabs.harvard.edu/abs/2011JMLR...12.2825P} {12, 2825}

\bibitem[\protect\citeauthoryear{Pelisoli et~al.,}{Pelisoli et~al.}{2023}]{Pelisoli2023}
Pelisoli I.,  et~al., 2023, \mn@doi [Nature Astronomy] {10.1038/s41550-023-01995-x}, 7, 931

\bibitem[\protect\citeauthoryear{Peters}{Peters}{2023}]{Peters2023}
Peters N.,  2023, The first transient search with automatically created MeerKAT images.
MSc Thesis, University of Amsterdam, \url {https://scripties.uba.uva.nl/search?id=record_54068}

\bibitem[\protect\citeauthoryear{Pritchard et~al.,}{Pritchard et~al.}{2021}]{Pritchard2021}
Pritchard J.,  et~al., 2021, \mn@doi [\mnras] {10.1093/mnras/stab299}, 502, 5438

\bibitem[\protect\citeauthoryear{{Pritchard}, {Murphy}, {Heald}, {Wheatland}, {Kaplan}, {Lenc}, {O'Brien}  \& {Wang}}{{Pritchard} et~al.}{2024}]{Pritchard2023}
{Pritchard} J.,  {Murphy} T.,  {Heald} G.,  {Wheatland} M.~S.,  {Kaplan} D.~L.,  {Lenc} E.,  {O'Brien} A.,   {Wang} Z.,  2024, \mn@doi [\mnras] {10.1093/mnras/stae127}, \href {https://ui.adsabs.harvard.edu/abs/2024MNRAS.529.1258P} {529, 1258}

\bibitem[\protect\citeauthoryear{{Protopapas}, {Huijse}, {Est{\'e}vez}, {Zegers}, {Pr{\'\i}ncipe}  \& {Marquette}}{{Protopapas} et~al.}{2015}]{Protopapas2015}
{Protopapas} P.,  {Huijse} P.,  {Est{\'e}vez} P.~A.,  {Zegers} P.,  {Pr{\'\i}ncipe} J.~C.,   {Marquette} J.-B.,  2015, \mn@doi [The Astrophysical Journal Supplement] {10.1088/0067-0049/216/2/25}, 216, 25

\bibitem[\protect\citeauthoryear{Pruzhinskaya, Malanchev, Kornilov, Ishida, Mondon, Volnova  \& Korolev}{Pruzhinskaya et~al.}{2019}]{Pruzhinskaya2019}
Pruzhinskaya M.~V.,  Malanchev K.~L.,  Kornilov M.~V.,  Ishida E.~E.,  Mondon F.,  Volnova A.~A.,   Korolev V.~S.,  2019, \mn@doi [\mnras] {10.1093/MNRAS/STZ2362}, 489, 3591

\bibitem[\protect\citeauthoryear{Pruzhinskaya et~al.,}{Pruzhinskaya et~al.}{2023}]{Pruzhinskaya2023}
Pruzhinskaya M.~V.,  et~al., 2023, \mn@doi [\aap] {10.1051/0004-6361/202245172}, 672, A111

\bibitem[\protect\citeauthoryear{{Rasmussen} \& {Williams}}{{Rasmussen} \& {Williams}}{2006}]{Rasmussen2006}
{Rasmussen} C.~E.,  {Williams} C. K.~I.,  2006, {Gaussian Processes for Machine Learning}.
The MIT Press, Cambridge

\bibitem[\protect\citeauthoryear{Rhodes, Caleb, Stappers, Andersson, Bezuidenhout, Driessen  \& Heywood}{Rhodes et~al.}{2023}]{Rhodes2023}
Rhodes L.,  Caleb M.,  Stappers B.~W.,  Andersson A.,  Bezuidenhout M.~C.,  Driessen L.~N.,   Heywood I.,  2023, \mn@doi [\mnras] {10.1093/mnras/stad2438}, 525, 3626

\bibitem[\protect\citeauthoryear{Richards et~al.,}{Richards et~al.}{2011}]{Richards2011}
Richards J.~W.,  et~al., 2011, \mn@doi [\apj] {10.1088/0004-637X/733/1/10}, 733, 10

\bibitem[\protect\citeauthoryear{{Rogers}, {Lintott}, {Croft}, {Schwamb}  \& {Davenport}}{{Rogers} et~al.}{2024}]{Rogers2024}
{Rogers} B.,  {Lintott} C.~J.,  {Croft} S.,  {Schwamb} M.~E.,   {Davenport} J. R.~A.,  2024, \mn@doi [\aj] {10.3847/1538-3881/ad1f5a}, \href {https://ui.adsabs.harvard.edu/abs/2024AJ....167..118R} {167, 118}

\bibitem[\protect\citeauthoryear{{Rowlinson} et~al.,}{{Rowlinson} et~al.}{2019}]{Rowlinson2019A&C....27..111R}
{Rowlinson} A.,  et~al., 2019, \mn@doi [Astronomy and Computing] {10.1016/j.ascom.2019.03.003}, \href {https://ui.adsabs.harvard.edu/abs/2019A&C....27..111R} {27, 111}

\bibitem[\protect\citeauthoryear{Rowlinson et~al.,}{Rowlinson et~al.}{2022}]{Rowlinson2022}
Rowlinson A.,  et~al., 2022, \mn@doi [\mnras] {10.1093/mnras/stac2460}, 517, 2894

\bibitem[\protect\citeauthoryear{Sainburg, McInnes  \& Gentner}{Sainburg et~al.}{2021}]{sainburg202UMAP}
Sainburg T.,  McInnes L.,   Gentner T.~Q.,  2021, Neural Computation, 33, 2881

\bibitem[\protect\citeauthoryear{{Sarbadhicary} et~al.,}{{Sarbadhicary} et~al.}{2021}]{Sarbadhicary2021}
{Sarbadhicary} S.~K.,  et~al., 2021, \mn@doi [\apj] {10.3847/1538-4357/ac2239}, \href {https://ui.adsabs.harvard.edu/abs/2021ApJ...923...31S} {923, 31}

\bibitem[\protect\citeauthoryear{Scargle}{Scargle}{1982}]{Scargle1982}
Scargle J.~D.,  1982, \mn@doi [\apj] {10.1086/160554}, 263, 835

\bibitem[\protect\citeauthoryear{{Smirnov} et~al.,}{{Smirnov} et~al.}{2024}]{Smirnov2024}
{Smirnov} O.~M.,  et~al., 2024, \mn@doi [\mnras] {10.1093/mnras/stae303}, \href {https://ui.adsabs.harvard.edu/abs/2024MNRAS.tmp..430S} {}

\bibitem[\protect\citeauthoryear{{Smirnov} et~al.,}{{Smirnov} et~al.}{2025}]{Smirnov2025}
{Smirnov} O.~M.,  et~al., 2025, \mn@doi [\mnras] {10.1093/mnrasl/slaf015}, \href {https://ui.adsabs.harvard.edu/abs/2025MNRAS.tmpL..14S} {}

\bibitem[\protect\citeauthoryear{Sooknunan et~al.,}{Sooknunan et~al.}{2021}]{Sooknunan2021}
Sooknunan K.,  et~al., 2021, \mn@doi [\mnras] {10.1093/mnras/staa3873}, 502, 206

\bibitem[\protect\citeauthoryear{{Stappers}}{{Stappers}}{2016}]{Stappers2016}
{Stappers} B.,  2016, in MeerKAT Science: On the Pathway to the SKA. p.~10, \mn@doi{10.22323/1.277.0010}

\bibitem[\protect\citeauthoryear{Swinbank et~al.,}{Swinbank et~al.}{2015}]{Swinbank2015}
Swinbank J.~D.,  et~al., 2015, \mn@doi [Astronomy and Computing] {10.1016/j.ascom.2015.03.002}, 11, 25

\bibitem[\protect\citeauthoryear{{The Astropy Collaboration} et~al.,}{{The Astropy Collaboration} et~al.}{2013}]{Astropy2013}
{The Astropy Collaboration} et~al., 2013, \mn@doi [\aap] {10.1051/0004-6361/201322068}, 558, A33

\bibitem[\protect\citeauthoryear{{The Astropy Collaboration} et~al.,}{{The Astropy Collaboration} et~al.}{2018}]{Astropy2018}
{The Astropy Collaboration} et~al., 2018, \mn@doi [\aj] {10.3847/1538-3881/aabc4f}, 156, 123

\bibitem[\protect\citeauthoryear{{The Astropy Collaboration} et~al.,}{{The Astropy Collaboration} et~al.}{2022}]{Astropy2022}
{The Astropy Collaboration} et~al., 2022, \mn@doi [\apj] {10.3847/1538-4357/ac7c74}, 935, 167

\bibitem[\protect\citeauthoryear{Villar, Cranmer, Berger, Contardo, Ho, Hosseinzadeh  \& Lin}{Villar et~al.}{2021}]{Villar2021}
Villar V.~A.,  Cranmer M.,  Berger E.,  Contardo G.,  Ho S.,  Hosseinzadeh G.,   Lin J. Y.-Y.,  2021, \mn@doi [\apjs] {10.3847/1538-4365/ac0893}, 255, 24

\bibitem[\protect\citeauthoryear{Virtanen et~al.,}{Virtanen et~al.}{2020}]{2020SciPy-NMeth}
Virtanen P.,  et~al., 2020, \mn@doi [Nature Methods] {10.1038/s41592-019-0686-2}, \href {https://rdcu.be/b08Wh} {17, 261}

\bibitem[\protect\citeauthoryear{Walmsley et~al.,}{Walmsley et~al.}{2020}]{Walmsley2020MNRAS.491.1554W}
Walmsley M.,  et~al., 2020, \mn@doi [\mnras] {10.1093/mnras/stz2816}, \href {https://ui.adsabs.harvard.edu/abs/2020MNRAS.491.1554W} {491, 1554}

\bibitem[\protect\citeauthoryear{Walmsley et~al.,}{Walmsley et~al.}{2022a}]{Walmsley2022a}
Walmsley M.,  et~al., 2022a, \mn@doi [\mnras] {10.1093/mnras/stab2093}, \href {https://ui.adsabs.harvard.edu/abs/2022MNRAS.509.3966W} {509, 3966}

\bibitem[\protect\citeauthoryear{Walmsley et~al.,}{Walmsley et~al.}{2022b}]{Walmsley2022b}
Walmsley M.,  et~al., 2022b, \mn@doi [\mnras] {10.1093/mnras/stac525}, 513, 1581

\bibitem[\protect\citeauthoryear{Wang, Tuntsov, Murphy, Lenc, Walker, Bannister, Kaplan  \& Mahony}{Wang et~al.}{2021a}]{YWang2021}
Wang Y.,  Tuntsov A.,  Murphy T.,  Lenc E.,  Walker M.,  Bannister K.,  Kaplan D.~L.,   Mahony E.~K.,  2021a, MNRAS, 502, 3294

\bibitem[\protect\citeauthoryear{Wang et~al.,}{Wang et~al.}{2021b}]{ZWang2021}
Wang Z.,  et~al., 2021b, \mn@doi [\apj] {10.3847/1538-4357/ac2360}, 920, 45

\bibitem[\protect\citeauthoryear{Wang et~al.,}{Wang et~al.}{2022}]{ZWang2022}
Wang Z.,  et~al., 2022, \mn@doi [\mnras] {10.1093/mnras/stac2542}, 516, 5972

\bibitem[\protect\citeauthoryear{Wang et~al.,}{Wang et~al.}{2023}]{YWang2023}
Wang Y.,  et~al., 2023, \mn@doi [\mnras] {10.1093/mnras/stad1727}, 523, 5661

\bibitem[\protect\citeauthoryear{Webb et~al.,}{Webb et~al.}{2020}]{Webb2020}
Webb S.,  et~al., 2020, \mn@doi [\mnras] {10.1093/mnras/staa2395}, 498, 3077

\bibitem[\protect\citeauthoryear{Webb et~al.,}{Webb et~al.}{2021}]{Webb2021}
Webb S.,  et~al., 2021, \mn@doi [\mnras] {10.1093/mnras/stab1798}, 506, 2089

\bibitem[\protect\citeauthoryear{Wright et~al.,}{Wright et~al.}{2017}]{Wright2017}
Wright D.~E.,  et~al., 2017, \mn@doi [\mnras] {10.1093/MNRAS/STX1812}, 472, 1315

\bibitem[\protect\citeauthoryear{{de Ruiter} et~al.,}{{de Ruiter} et~al.}{2024}]{deRuiter2024arXiv240811536D}
{de Ruiter} I.,  et~al., 2024, \mn@doi [arXiv e-prints] {10.48550/arXiv.2408.11536}, \href {https://ui.adsabs.harvard.edu/abs/2024arXiv240811536D} {p. arXiv:2408.11536}

\makeatother
\end{thebibliography}







\bsp	
\label{lastpage}
\end{document}